\documentclass[12pt]{iopart}

\usepackage{graphicx}
\usepackage{epstopdf}
\usepackage{float}
\usepackage{soul}

\usepackage{iopams}
\usepackage{setstack}
\usepackage{comment}
\usepackage[T1]{fontenc}

\usepackage{color}
\definecolor{aurometalsaurus}{rgb}{0.43, 0.5, 0.5}

\renewcommand{\text}[1]{}

\usepackage[sep=7.5pt, offset=0.6em]{simpler-wick}

\usepackage{citesort}

\begin{document}

\title[Realistic estimates of superconducting properties for the cuprates \ldots]{Realistic estimates of superconducting properties for the cuprates: reciprocal-space diagrammatic expansion combined with variational approach}

\author{M. Fidrysiak${}^1$, M. Zegrodnik${}^2$, and J. Spa{\l}ek${}^1$}
\address{${}^1$Marian Smoluchowski Institute of Physics, Jagiellonian University, ul. {\L}ojasiewicza 11, 30-348 Krak\'{o}w, Poland}
\address{${}^2$Academic Centre for Materials and Nanotechnology, AGH University of Science and Technology,
Al. Mickiewicza 30, 30-059 Krak\'{o}w, Poland}

\vspace{10pt}

\begin{abstract}
We propose a systematic approach to the systems of correlated electrons, the so-called $\mathbf{k}$-DE-GWF method, based on reciprocal-space ($\mathbf{k}$-resolved) diagrammatic expansion of the variational Gutzwiller-type wave function for parametrized models of correlated fermions. The present approach, in contrast to either variational Monte-Carlo (VMC), or the recently developed real-space diagrammatic expansion of the Gutzwiller-type wave function (direct-space DE-GWF technique), is applicable directly in the thermodynamic limit and thus is suitable for describing selected singular features of the wave-vector-dependent quantities. We employ the $\mathbf{k}$-DE-GWF method to extract the non-analytic part of the two leading moments of the fermion spectral-density function across the (two-dimensional) Brillouin zone for the Hubbard model and away from the half-filling. Those moments are used to evaluate the  nodal quasiparticle velocities and their spectral weights in the correlated superconducting state. The two velocities determined in that manner exhibit scaling with the electron concentration qualitatively different from that obtained earlier for the excited states of the high-$T_c$ cuprates within the projected quasi-particle ansatz, and the results are in a very good quantitative agreement with experimental data if interpreted as those characterizing the spectrum below and above the observed kink. We provide a detailed discussion of the two gaps and two excitation branches (two velocities) appearing naturally within our DE-GWF approach. The two separate sets of characteristics distinguish the renormalized quasiparticle states very close to the Fermi surface from the deeper correlated-state properties. Also, an enhancement of the $\mathbf{k}$-dependent magnetic susceptibility is shown to contain a spin-fluctuation contribution within our language. Finally, the $\mathbf{k}$-DE-GWF approach is compared to both the VMC and real-space DE-GWF results for the cases of Hubbard and $t$-$J$-$U$ models. 
\end{abstract}

%
%
%
%
%

\section{Introduction}

Devising the theory of strongly correlated fermions is one of the most longstanding and challenging problems of condensed matter physics, both in the case of electronic and cold-atom lattice systems of fermions. The difficulty is caused by the circumstance that, in such a many-particle system, the mutual interaction is comparable, or can even exceed by far, the kinetic- or band-energy contribution to the total energy. In such a situation, the standard perturbation treatment of the interaction part is not admissible, since the starting point of this many-fermion state is neither that of electron gas nor Landau-Fermi-liquid \cite{BookPines,BookBaym}. Instead, the Mott-Hubbard insulating phase of localized (quasi-atomic) states that sets in, as has been elaborated first by Mott \cite{BookMott}, Anderson \cite{BookAnderson}, and Hubbard \cite{HubbardProcRoyalPhysSoc1964}. In effect, the correlated systems undergo a discontinuous insulator-metal (localized-delocalized) phase transition \cite{ImadaRevModPhys1998,HonigChemMater1998} and one has to incorporate such limiting situations as a transformation between the two complementary, momentum and position, quantum-mechanical representations. Also, one has to encompass the Hartree-Fock (weak-interaction) and the kinetic-exchange (strong-correlation) regimes as the opposite asymptotic limits. The interest in these problems acquired an additional dimension with the discovery of the Mott-insulating state evolution into the high-temperature superconducting phase (high-$T_c$ SC) \cite{OgataRepProgPhys2008,BookRanderia}, as well as the superfluid-Mott insulator transition in the cold-atom systems in optical lattices \cite{ZwergerJOptB2003,BookLewenstein}. The latter aspects point to a universal connection between localization-delocalization (Mott-Hubbard) transition and SC (superfluidity) in the correlated fermionic and bosonic systems. The question that still remains is the  relation between the Mott transition and the superconductivity (superfluidity). In particular,  whether the mechanism of the superconducting pairing can be traced back to specific features of the insulating state that become operative once the metallic phase stabilizes.

Due to these complications, two main approaches have been developed. One of them starts from microscopic, but parametrized models such as Hubbard, Anderson-lattice, or $t$-$J$-($U$) models, and when solving them, the dominant nature of interelectronic correlations is emphasized. Within this approach, the exact solutions have been obtained for the special cases of one-dimensional lattice \cite{LiebPhysRevLett1968,LiebBook2005}. In the spatial dimension $d \geq 1$, advanced numerical methods based on either quantum Monte-Carlo \cite{BeccaBook2017}, cluster expansion, or renormalization group have been implemented for finite (if not small) systems, sometimes with tightly connected finite-size-scaling analysis \cite{LeBlancPhysRevX2015}, to extend their validity to the properties of extended systems.

A separate class of solutions is based on the variational approach by selecting either Gutzwiller \cite{MetznerPhysRevB1988,BunemannEPL2012} or Jastrow \cite{BaeriswylChapter1987} type of the many-particle wave function. The variational approach, developed systematically, allows for a detailed analysis of high-$T_c$ superconducting systems in the thermodynamic limit \cite{KaczmarczykPhysRevB2013}, and one specific version of it is the subject of the present paper. An alternative approach is based on incorporation of correlations in advanced band-theoretical  calculations, which are based on the density functional theory (DFT). In this class we have the LDA+$U$ \cite{HaulePhysRevLett2015,ChenPhysRevB2016}, LDA+DMFT \cite{VollhardtDMFT2014,KotliarRevModPhys2006}, or LDA+Gutzwiller \cite{TianPhysRevB2015,SchicklingNewJPhys2014} methods. These methods that mix an \textit{ab initio} and the parametrized-model approaches, have been used quite frequently and are applied to concrete materials. Nonetheless, one has to keep in mind that an inherent problem of all these approaches methodologies is double counting of interparticle interaction, overcoming of which is under debate. Also, they mix both \textit{ab initio} and parametrized-model aspects of the problem. In connection with this, we can also mention our earlier work of combining the \textbf{E}xact \textbf{D}iagonalization and \textbf{Ab} \textbf{I}nitio approach (EDABI method), which is free of double counting problem, as it does not reintroduce the interaction parameters into the \textit{ab initio} calculation scheme. However, so far this method can be implemented effectively only to either finite-size systems \cite{BiborskiSciRep2016}, or to the model systems with a small number of valence orbitals per atom \cite{SpalekJPhysCondensMatter2007}.

Here we discuss  a variant of the variational approach for parametrized models, that is based on a systematic expansion of the Gutzwiller wave function, this time in its reciprocal space version. The previous, real-space treatment \cite{KaczmarczykNewJPhys2014,SpalekPhysRevB2017_95_024506,ZegrodnikPhysRevB2017_95_024507,ZegrodnikPhysRevB2017,WysokinskiPhysRevB2016} allowed us to include correlations in real-space extending to few lattice constants. That approach contains, apart from its advantages, also some inaccuracies when considering explicitly the wave vector, $\mathbf{k}$-dependent, quantities such as the statistical distribution function or the spectral-density in the quasiparticle terms for  correlated metallic state. Within the present method we define first the quasiparicle characteristics. Second, we obtain the related quantities such as the Fermi velocities or the $\mathbf{k}$-dependent magnetic susceptibility enhancement, which are evaluated explicitly in the two dimensional situation, considered explicitly in the paper. In our view, this analysis proposes  a substantial progress in the sense that the $\mathbf{k}$-dependent quantities can be obtained for extended (infinite) systems in a systematic manner. Also, the approach can be extended to a wide class of correlated models, not limited to those discussed below.

The paper is organized as follows. In section~\ref{sec:method} we describe the $\mathbf{k}$-DE-GWF method. In section~\ref{sec:hubbard_model} we apply it to the Hubbard model. Specifically, in section~\ref{sec:k_de_gwf_solution} the variational solution as a function of hole-doping is discussed. In section~\ref{sec:spectral_functions} we report the results for the two leading spectral function moments and extract the two Fermi velocities and the corresponding weights. In section~\ref{sec:comparison_experiment} and \ref{sec:methodological_remark} we compare our results with experiment and provide additional methodological remarks, respectively. Section \ref{sec:qp_anisotropy} details the correlation-induced anisotropy of the quasiparticle properties. Finally, in section~\ref{sec:spin_correlations} we provide the analysis of instantaneous spin correlations and compare the results with the available VMC data. In section~\ref{sec:discussion} we overview the results and discuss briefly the relation and difference between the Hubbard or $t$-$J$ models, as well as our earlier results for the $t$-$J$-$U$ model. We also make a suggestion there about a possible relation between our effective gap and the pseudogap since two separate energy scales appear in the present formulation, but this particular point requires a further future analysis. Nontrivial technical details of the analysis are transferred to Appendices A-D.

\section{Reciprocal-space diagrammatic expansion for the Gutzwiller wave function: the $\mathbf{k}$-DE-GWF method}
\label{sec:method}

\subsection{Real-space diagrammatic expansion: A brief summary}

For the reader's convenience, we first summarize the previously developed \cite{BunemannEPL2012,KaczmarczykNewJPhys2014} real-space diagrammatic expansion of the Gutzwiller wave function (DE-GWF) which constitutes the starting point for our  $\mathbf{k}$-DE-GWF approach. In the subsequent discussion we consider the fairly general $t$-$J$-$U$ model, within which the Hubbard and $t$-$J$ models can be regarded formally as particular cases \cite{SpalekPhysRevB2017_95_024506}. The $t$-$J$-$U$ model is defined by the parametrized Hamiltonian

\begin{eqnarray}
\mathcal{H} \ = \sum \limits_{ij\sigma} t_{ij} c^\dagger_{i \sigma} c_{j \sigma} + U \sum \limits_{i} \hat{n}_{i\uparrow} \hat{n}_{i \downarrow} + J \sum \limits_{\langle i, j\rangle} \mathbf{\hat{S}}_i \cdot \mathbf{\hat{S}}_j. \label{eq:tju-model}
\end{eqnarray}

\noindent
In this expression, the first term is the hopping (kinetic-energy) part, here with nonzero magnitudes $t < 0$ and $t' = 0.25 |t| > 0$ for the nearest and next-nearest neighbors, respectively. The second denotes the intraatomic Coulomb interaction $\sim U \gg |t|$, and the third is the spin-dependent antiferromagnetic interaction with the dominant nearest-neighbor integral $J$ (the symbol $\langle i, j\rangle$ indicates summation over pairs of nearest neighbors). The particular cases are: the Hubbard model for $J = 0$ and the $t$-$J$ model for $J > 0$ and $U \gg |t|$. Most of the further discussion is focused on the Hubbard-model ($J \rightarrow 0$) limit. Nonetheless, calculations of limited scope for  $J > 0$ have been preformed to make comparison with VMC and our earlier real-space DE-GWF results. The applicability of this model to high-$T_c$ cuprates has been discussed elsewhere \cite{SpalekPhysRevB2017_95_024506,ZegrodnikPhysRevB2017_95_024507,ZegrodnikPhysRevB2017,AbramJPCM2017}.

The Gutzwiller wave function method is based on minimization of the energy functional $E_G = \left<\Psi_G| \mathcal{H} |\Psi_G\right> / \left<\Psi_G|\Psi_G\right>$, with the trial state $|\Psi_G\rangle = P_G |\Psi_0\rangle$, where $|\Psi_0\rangle$ is a wave function of uncorrelated fermions, to be defined later as a state with broken symmetry. The operator $P_G = \prod_i P_{Gi}$ changes the relative weights of local many-particle states $|0\rangle_i$, $|\uparrow\rangle_i$, $|\downarrow\rangle_i$, and $|\uparrow\downarrow\rangle_i$ on lattice site $i$, namely $P_{Gi} = \lambda_0 |0\rangle_i {}_i\langle0| + \lambda_\uparrow |\uparrow\rangle_i {}_i\langle\uparrow| + \lambda_\downarrow |\downarrow\rangle_i {}_i\langle\downarrow| + \lambda_{\uparrow\downarrow} |\uparrow\downarrow \rangle_i {}_i\langle\uparrow\downarrow|$ \cite{BunemannEPL2012}. The normalization factor $\left<\Psi_G|\Psi_G\right>$  needs to be introduced, since $P_G$ is not a unitary operator. For the sake of compactness, we adopt the notation $\langle \hat{\mathcal{O}} \rangle_G \equiv \langle \Psi_G| \hat{\mathcal{O}} |\Psi_G\rangle /\langle\Psi_G|\Psi_G\rangle$ for the expectation  values of any operator $\hat{\mathcal{O}}$.

From now on, we assume the spin-rotational and lattice-translational symmetries, which simplifies substantially the discussion. In that situation, the kinetic-, Hubbard-, and exchange-contributions to the variational energy, i.e.,  $E_G \equiv E_\mathrm{kin} + E_U + E_J$, can be written as a sum of three respective terms

\begin{eqnarray}
  &E_\mathrm{kin}  \equiv \sum\limits_{ij\sigma} t_{ij} \langle c^\dagger_{i\sigma} c_{j\sigma}\rangle_G  =  \frac{\sum \limits_{ij} t_{ij} \langle \Psi_0| P_{Gi}c^\dagger_{i\sigma}P_{Gi} P_{Gj}c_{j\sigma} P_{Gj} \prod\limits_{l\neq i, j} P_{Gl}|\Psi_0 \rangle}{\langle\Psi_G|\Psi_G\rangle},\label{eq:kinetic_energy_contibution} \\
  &E_U \equiv U \sum\limits_{i} \langle \hat{n}_{i\uparrow}\hat{n}_{i\downarrow} \rangle_G  =  \frac{U \sum\limits_i \langle \Psi_0| P_{Gi}\hat{n}_{i\uparrow}\hat{n}_{i\downarrow}P_{Gi} \prod \limits_{l\neq i} P_{Gl}|\Psi_0 \rangle}{\langle\Psi_G|\Psi_G\rangle}, \label{eq:hubbard_energy_contibution}
\end{eqnarray}

\noindent
and

\begin{eqnarray}
  E_J \equiv &\frac{3}{4} J \sum \limits_{\langle i, j \rangle}  \langle \hat{S}^+_i \hat{S}^-_j \rangle_G + \mathrm{c.c.} = \nonumber\\&= \frac{\frac{3}{4} J \sum\limits_{\langle i, j \rangle}\langle \Psi_0| P_{Gi}\hat{S}^{+}_{i}P_{Gi} P_{Gj}\hat{S}^{-}_{j} P_{Gj} \prod \limits_{l\neq i, j} P_{Gl}|\Psi_0 \rangle}{\langle\Psi_G|\Psi_G\rangle} + \mathrm{c.c.},
\label{eq:exchange_energy_contibution}
\end{eqnarray}

\noindent
where $\hat{S}^{+}_i \equiv c^\dagger_{i\uparrow} c_{i\downarrow}$ and $\hat{S}^{-}_i \equiv c^\dagger_{i\downarrow} c_{i\uparrow}$ are spin operators. Note that in the exchange part $E_J$ it is sufficient to consider the transverse component $\hat{S}^+_i\hat{S}^-_j$ due to the assumed spin-rotational symmetry in pure SC or paramagnetic states, the only states analyzed here.

Evaluation of the above expectation values using the Gutzwiller-correlated wave function is a non-trivial many-particle problem that, for finite lattices, can be handled, e.g., by variational Monte-Carlo (VMC) approach \cite{CeperleyPhysRevB1977,FoulkesRevModPhys2001}. Here we follow a different path and employ Wick's theorem to evaluate them in real space. This is possible since $|\Psi_0\rangle$ does not contain directly intersite correlations, though it will be selected as a broken-symmetry (SC) state optimized variationally as well. To rationalize the resultant diagrammatic decomposition with a formal expansion parameter, an additional constraint $P_{Gi}^2 = 1 + x \times \hat{d}_i^\mathrm{HF}$ needs to be imposed \cite{BunemannEPL2012}, with $\hat{d}_i^\mathrm{HF} \equiv (\hat{n}_{i\uparrow} - n^0_{\uparrow}) (\hat{n}_{i\downarrow} - n^0_{\downarrow})$ and $n^0_\sigma \equiv \langle\Psi_0|\hat{n}_{i\sigma}|\Psi_0\rangle$. The latter formal trick allows to express all four correlator coefficients $\{\lambda_{\alpha}\}$ in terms of a single variational parameter $x$. Explicitly, $\lambda_0^2 = 1 + x  n^0_\uparrow n^0_\downarrow$, $\lambda_\sigma^2 = 1 - x n^0_\sigma (1 - n^0_{\bar{\sigma}})$, and $\lambda_{\uparrow\downarrow}^2 = 1 + x (1 - n^0_\sigma) (1 - n^0_{\bar{\sigma}})$, with $\sigma \equiv \uparrow$ or $\downarrow$. By evaluating the on-site operator products $P_{Gi} c_{i\sigma} P_{Gi}$, $P_{Gi} \hat{n}_{i\sigma} P_{Gi}$, $P_{Gi} \hat{S}^+_i P_{Gi}$, and  $P_{Gi} \hat{n}_{i\uparrow}\hat{n}_{i\downarrow} P_{Gi}$, and Taylor-expanding the remaining terms in powers of parameter $x$, all the above quantities can be rewritten in terms of closed-form  expressions, namely

\begin{eqnarray}
  \label{eq:kinetic_energy_contributio_diag}
  E_\mathrm{kin} = 2 \times \sum \limits_{ij} t_{ij} (q^2 T^{11}_{ij} + 2 q \alpha T^{13}_{ij} + \alpha^2 T^{33}_{ij}),
\end{eqnarray}
\begin{eqnarray}
  \label{eq:hubbard_energy_contributio_diag}
  E_U = \gamma^{(4)} I^4 + \gamma^{(2)} I^2 + \lambda_{\uparrow\downarrow}^2 n^0_\uparrow n^0_\downarrow,
\end{eqnarray}
\noindent
and
\begin{eqnarray}
  \label{eq:exchange_energy_contributio_diag}
  E_J = \frac{3}{4} J \lambda_{\uparrow}^2 \lambda_{\downarrow}^2 \sum \limits_{\langle i, j \rangle} S^{22}_{ij} + \mathrm{c.c},
\end{eqnarray}

\noindent
where

\begin{eqnarray}
  T^{11}_{ij} &= \sum \limits_{k=0}^{\infty} \frac{x^k}{k!} \sum_{l_1, \ldots, l_k} {}^{'} \langle\Psi_0| c^\dagger_{i\uparrow} c_{j\uparrow} \hat{d}^\mathrm{HF}_{l_1} \ldots \hat{d}^\mathrm{HF}_{l_k}|\Psi_0\rangle_c, \label{eq:t11} \\
  T^{13}_{ij} &= \sum \limits_{k=0}^{\infty} \frac{x^k}{k!} \sum_{l_1, \ldots, l_k}{}^{'} \langle\Psi_0| c^\dagger_{i\uparrow} c_{j\uparrow} \hat{n}^\mathrm{HF}_{j\downarrow} \hat{d}^\mathrm{HF}_{l_1} \ldots \hat{d}^\mathrm{HF}_{l_k}|\Psi_0\rangle_c, \\
  T^{33}_{ij} &= \sum \limits_{k=0}^{\infty} \frac{x^k}{k!} \sum_{l_1, \ldots, l_k}{}^{'} \langle\Psi_0| c^\dagger_{i\uparrow}\hat{n}^\mathrm{HF}_{i\downarrow} c_{j\uparrow} \hat{n}^\mathrm{HF}_{j\downarrow} \hat{d}^\mathrm{HF}_{l_1} \ldots \hat{d}^\mathrm{HF}_{l_k}|\Psi_0\rangle_c, \label{eq:T33_definition}\\
  S^{22}_{ij} &= \sum \limits_{k=0}^{\infty} \frac{x^k}{k!}  \sum_{l_1, \ldots, l_k}{}^{'} \langle\Psi_0| \hat{S}^{+}_i \hat{S}^{-}_j \hat{d}^\mathrm{HF}_{l_1} \ldots \hat{d}^\mathrm{HF}_{l_k}|\Psi_0\rangle_c, \\
  I^{2} &= \sum \limits_{k=0}^{\infty} \frac{x^k}{k!} \sum_{l_1, \ldots, l_k}{}^{'} \langle\Psi_0| \hat{n}_{i\sigma}^\mathrm{HF} \hat{d}^\mathrm{HF}_{l_1} \ldots \hat{d}^\mathrm{HF}_{l_k}|\Psi_0\rangle_c, \\
  I^{4} &= \sum \limits_{k=0}^{\infty} \frac{x^k}{k!}  \sum_{l_1, \ldots, l_k}{}^{'}  \langle\Psi_0| \hat{d}_i^\mathrm{HF} \hat{d}^\mathrm{HF}_{l_1} \ldots \hat{d}^\mathrm{HF}_{l_k}|\Psi_0\rangle_c. \label{eq:i4}
\end{eqnarray}

\noindent
The  subscript ``$c$'' indicates that only the connected diagrams are included in the Wick's decomposition: The disconnected part is canceled out by the power expansion of the normalization factor $\langle\Psi_G|\Psi_G\rangle$. The indices $l_1, \ldots, l_k$ run over lattice sites, but with the restriction that all $l_n$, $i$, and $j$ must be different (this is indicated by primes next to the second summation symbols). Moreover, renormalization factors $q \equiv \lambda_\sigma \lambda_0 + n^0_{\bar{\sigma}} (\lambda_{\uparrow\downarrow}\lambda_{\bar{\sigma}} - \lambda_\sigma \lambda_0)$, $\alpha \equiv \lambda_{\uparrow\downarrow}\lambda_{\bar{\sigma}} - \lambda_\sigma \lambda_0$, $\gamma^{(2)} \equiv 2\lambda_{\uparrow\downarrow}^2 n^0_\sigma$, and $\gamma^{(4)} \equiv \lambda_{\uparrow\downarrow}^2 (1 - x n^0_\uparrow n^0_\downarrow)$ emerge due to Gutzwiller-correlator presence (note that one does not need to specify the spin index $\sigma$ in the definitions of $q$, $\alpha$, and $\gamma^{(2)}$ because of the spin-rotational invariance). The multi-site averages appearing in equations~(\ref{eq:t11})-(\ref{eq:i4}) can be evaluated by means of Wick's theorem by decoupling them into pairs and, thereby, are expressed in terms of the two-point expectation values (lines) $P_{ij\sigma} \equiv \langle \Psi_0| c^\dagger_{i\sigma} c_{j\sigma} |\Psi_0\rangle$ and the anomalous amplitudes $S_{ij} \equiv \langle \Psi_0| c_{i\uparrow} c_{j\downarrow} | \Psi_0\rangle$.

The final form of the energy functional, used subsequently in the minimization procedure of $E_G$, reads

\begin{eqnarray}
  \label{eq:optimization_problem}
  &\mathcal{F}(\{P_{ij\sigma}\}, \{S_{ij}\}, x, \mu, {\lambda^P_{\sigma ij}, {\lambda^S_{\sigma ij}}, E, |\Psi_0\rangle}) \equiv \nonumber \\ \equiv &\Omega_G(\{P_{ij\sigma}\}, \{S_{ij}\}, x, \mu) - \sum \limits_{ij\sigma} \lambda^P_{ij\sigma} (P_{ij\sigma} - \langle\Psi_0| c^\dagger_{i\sigma} c_{j\sigma} |\Psi_0\rangle) - \nonumber \\ - &\sum \limits_{ij} \lambda^S_{ij} (S_{ij} - \langle\Psi_0| c_{i\uparrow} c_{j\downarrow} |\Psi_0\rangle) + \mu N_e - E (\langle\Psi_0|\Psi_0\rangle - 1),
\end{eqnarray}

\noindent
where $\Omega_G \equiv E_G - \mu \langle \hat{N} \rangle_G$. The Lagrange multipliers $\lambda^P_{ij\sigma}$ and $\lambda^S_{ij}$ ensure that the correlation functions, $P_{ij\sigma}$ and $S_{ij}$, represent appropriate expectation values, evaluated with the uncorrelated wave function $|\Psi_0\rangle$. Additionally, we have relaxed the constraint of fixed particle number and introduced the chemical potential term $\mu$ as an additional Lagrange multiplier to ensure that the average occupancy equals to $N_e$ \cite{SpalekJPhysCondensMatter2013}. Finally, the last constraint (with the Lagrange multiplier $E$) is to keep the normalization of the uncorrelated wave function, $\langle\Psi_0|\Psi_0\rangle = 1$.

The constrained optimization problem, given by functional~(\ref{eq:optimization_problem}), results in the system of equations

\begin{eqnarray}
  \lambda^P_{ij\sigma} = \frac{\partial \Omega_G}{\partial P_{ij\sigma}},  \lambda^S_{ij} = \frac{\partial \Omega_G}{\partial S_{ij}}, \label{eq:var_eq_begin}\\
  P_{ij\sigma} = \langle \Psi_0|c^\dagger_{i\sigma} c_{j\sigma}|\Psi_0 \rangle, S_{ij} = \langle \Psi_0|c_{i\uparrow} c_{j\downarrow}|\Psi_0 \rangle, \\
  \langle\hat{N}\rangle_G = N_e, \\
  \frac{\partial E_G}{\partial x} = 0,
\end{eqnarray}

\noindent
and, additionally,
  \begin{eqnarray}
  \mathcal{H}_\mathrm{eff} |\Psi_0\rangle = E |\Psi_0\rangle. \label{eq:var_eq_end}
\end{eqnarray}

\noindent

\begin{figure}
  \centering
\includegraphics[width = 0.45\columnwidth]{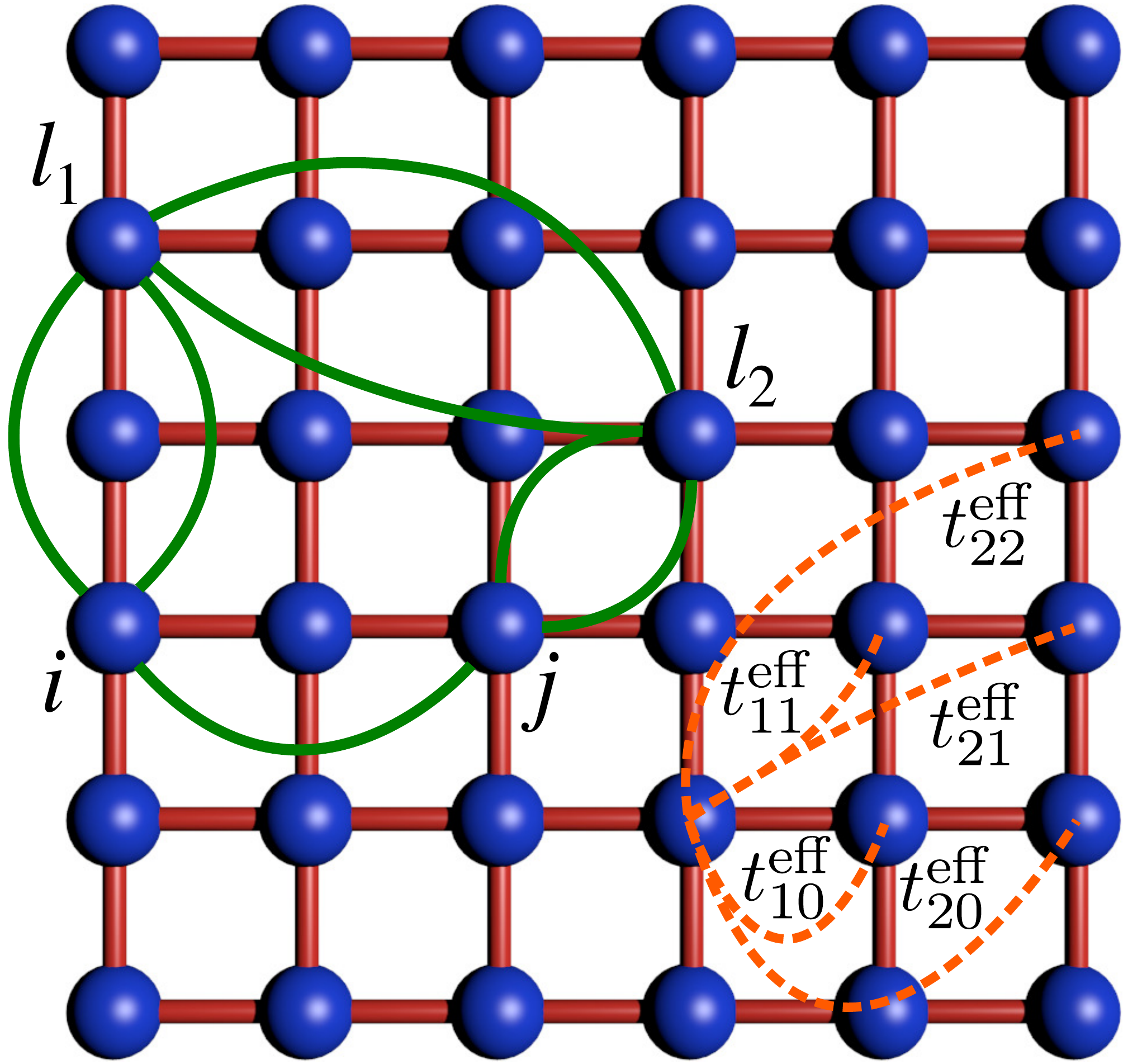} %
\caption{Illustration of the square lattice and selected real-space objects involved in the $\mathbf{k}$-DE-GWF procedure. Specifically, exemplary second-order ($k = 2$) real-space diagram contributing to the expectation value $T^{33}_{ij}$ is drawn by green lines representing the hopping amplitude $\langle\Psi_0| c^\dagger_{i\sigma} c_{j\sigma} |\Psi_0\rangle$ between external vertices $i$ and $j$ and virtual processes involving internal sites $l_1$ and $l_2$. The summation is performed over indices $l_1$ and $l_2$ (cf. equation~(\ref{eq:T33_definition})). The orange dashed lines represent a few shortest-range hopping parameters entering the effective Hamiltonian $\mathcal{H}_\mathrm{eff}$ (cf. equation~(\ref{eq:heff})).}%
  \label{fig:diagram_real_space}
\end{figure}

The eigenequation~(\ref{eq:var_eq_end}) is obtained from optimization of $\mathcal{F}$ with respect to uncorrelated wave function $\langle\Psi_0|$, i.e., $\mathcal{H}_\mathrm{eff} |\Psi_0\rangle = \frac{\delta}{\delta{\langle\Psi_0|}} \{\mathcal{F} + E\langle\Psi_0|\Psi_0\rangle\}$ so that equation~(\ref{eq:var_eq_end}) follows from $\frac{\partial \mathcal{F}}{\partial \langle\Psi_0|} = 0$. By implementing this equation, the starting wave function $|\Psi_0\rangle$ is also obtained variationally, i.e., in a self-consistent manner. This is an essential feature of the DE-GWF approach. Explicitly, the effective Hamiltonian has the form

\begin{eqnarray}
\label{eq:heff}
  \mathcal{H}_\mathrm{eff} = \sum \limits_{\overset{ij\sigma}{i \neq j}} t^\mathrm{eff}_{ij} c^\dagger_{i\sigma} c_{j\sigma} + \sum \limits_{ij} \left[ \Delta^\mathrm{eff}_{ij}  c_{i\uparrow} c_{j\downarrow} + \mathrm{H.c.} \right] - \mu^\mathrm{eff}\sum \limits_{i\sigma} \hat{n}_{i\sigma},
\end{eqnarray}

\noindent where $t^\mathrm{eff}_{ij} \equiv \frac{\partial \Omega_G}{\partial P_{ij\sigma}}$ and $\Delta^\mathrm{eff}_{ij} \equiv \frac{\partial \Omega_G}{\partial S_{ij}}$ denote the effective hopping integrals and pairing potential components, respectively. Additionally, the effective chemical potential $\mu^\mathrm{eff} \equiv - \frac{\partial \Omega_G}{\partial n^0_{i\sigma}}$ appears. A few shortest-range hopping integrals have been marked in figure~\ref{fig:diagram_real_space} by the orange dashed lines.

Note that $\mathcal{H}_\mathrm{eff}$ comes out formally as a supplemental entity after imposing the procedure of minimizing the ground-state energy also with respect the starting wave function $|\Psi_0\rangle$. Nonetheless, we argue in our later analysis that this additional feature of our approach can be given a precise physical interpretation. Namely, its spectrum (up to small, well controlled, corrections) corresponds to energies of the normalized projected quasi-particle states, defined as $|\Psi_\mathbf{k}\rangle \equiv P_G c^\dagger_\mathbf{k} | \Psi_\mathbf{k}\rangle || P_G c^\dagger_\mathbf{k} |\Psi_0\rangle||^{-1}$. Thus, $\mathcal{H}_\mathrm{eff}$ governs the dynamics of the projected quasiparticles which, as such, generalize the corresponding concept of quasiparticle state introduced in the context of renormalized mean-field theory (cf. \cite{EdeggerAdvPhys2007} and references therein). This statement is proved formally in \ref{appendix:effective_hamiltonian}.

\noindent

The principal technical difficulty of solving equations~(\ref{eq:var_eq_begin})-(\ref{eq:var_eq_end}) is in evaluating the Wick-decomposed functional $\Omega_G$ and its derivatives with respect to $P_{ij\sigma}$ and $S_{ij}$ (the latter are required to construct the effective Hamiltonian $\mathcal{H}_\mathrm{eff}$). This leads to a diagrammatic expansion, constituting the basis for the real-space \textbf{D}iagrammatic \textbf{E}xpansion of the \textbf{G}utzwiller \textbf{W}ave \textbf{F}unction (DE-GWF) method, elaborated in detail earlier \cite{KaczmarczykNewJPhys2014,SpalekPhysRevB2017_95_024506,ZegrodnikPhysRevB2017_95_024507,ZegrodnikPhysRevB2017} and successfully applied to the high-$T_c$ cuprates. In figure~\ref{fig:diagram_real_space} we show an exemplary second-order real-space DE-GWF graph (green lines) contributing to the diagrammatic sum $T^{33}_{ij}$.

\subsection{Reciprocal-space diagrammatic expansion $(\mathbf{k}$-$\mathrm{DE}$-$\mathrm{GWF})$}

\begin{figure}
\centering
\includegraphics[width = 0.65\columnwidth]{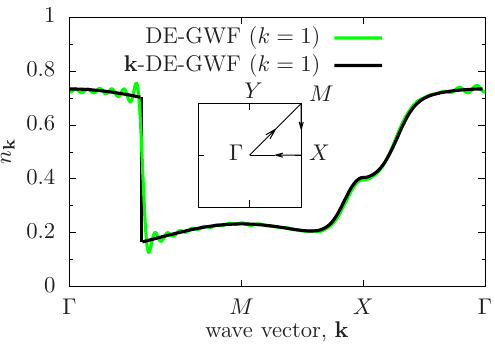} %
\caption{Statistical distribution function $n_\mathbf{k}$ in the correlated state $n_\mathbf{k}$ across the two-dimensional Brillouin-zone (BZ) contour $\Gamma$-$M$-$X$-$\Gamma$, evaluated up to the first order of the diagrammatic expansion ($k \leq 1$) within the DE-GWF approach and with the real-space cutoff of five lattice constants (green line), as well as by the new $\mathbf{k}$-DE-GWF method, developed here (black line). In the latter case, Cuba Suave algorithm was used for Monte-Carlo integration and $k \leq 1$ diagrams were included in the calculation. Note the Gibbs-type oscillations, apparent in the DE-GWF result, are absent in the $\mathbf{k}$-DE-GWF case, where there appears a clear discontinuity of $n_\mathbf{k}$ along the nodal ($\Gamma$-$M$) direction. Inset: the selected $\Gamma$-$M$-$X$-$\Gamma$ contour in the BZ. The model parameters are $t'/|t| = 0.25$, $U/|t| = 20$, $J/|t| = 1/3$, and the doping $\delta \approx 0.198$.}%
  \label{fig:comparison_mz}
\end{figure}

The procedure, described in the previous subsection, with the real-space diagrammatic expansion in the DE-GWF form, tends to converge rapidly with respect to the expansion order $k$ for $k > 2$. Nonetheless, the fundamental limitation of the this technique stems from the fact that the lattice-site summation over the internal vertex positions $\{l_i\}$ cannot be explicitly performed in the thermodynamic limit, since then the number of internal sites becomes infinite. An effective way to deal with this problem is then to introduce a cutoff for the lengths of the graph edges, typically three- to five- lattice constants. Such a limitation has proven to be irrelevant for the case of local quantities that are weakly influenced by the long-range correlation effects. However, as far as the non-local (wave-vector-resolved) quantities are concerned, the real-space cutoff smears out the physically meaningful discontinuities that appear, e.g., in the statistical distribution function $n_\mathbf{k}$, as well as induces other artifacts, such as the Gibbs-type oscillations. Consequently, the physical information encoded in those singularities is lost within this approach. To illustrate this point, in figure~\ref{fig:comparison_mz} we plot the statistical distribution function in the correlated state $n_\mathbf{k} \equiv \langle c^\dagger_{\mathbf{k}\sigma} c_{\mathbf{k}\sigma}\rangle_G$ for the $t$-$J$-$U$ model and calculated using the DE-GWF method in the $d$-wave SC state (green solid line) across the selected contour in the Brillouin zone; the values of parameters are provided in the plot caption. The plot encompasses the contour $\Gamma$-$M$-$X$-$\Gamma$ in the Brillouin zone (cf. the inset). The $d$-wave symmetry requires that the SC gap closes along the nodal $\Gamma$-$M$ direction so that a discontinuity of the distribution function is then expected. However, due to the real-space cutoff, inherent to the DE-GWF, only a steep albeit quasi-continuous character is achieved. The latter is accompanied by the so-called Gibbs oscillations. We point out that $n_\mathbf{k}$ can be also calculated using the VMC \cite{OtsukaJPhysSocJapan1992,TocchioPhysRevB2012}. However, due to the finite size of the system considered, VMC provides the values on a discrete $\mathbf{k}$-space mesh. The emerging non-analytic features of $n_\mathbf{k}$ and other $\mathbf{k}$-resolved quantities can thus be only estimated in a crude manner.

Here we propose an extension of the DE-GWF method that allows to eliminate the finite-range-summation artifacts and to account for the singular features of the wave vector resolved quantities in the correlated state by evaluating them  directly in the thermodynamic limit. The effect of such an extension for the case of the distribution function $n_\mathbf{k}$ is shown in figure~\ref{fig:comparison_mz} by a black solid line. In contrast to the DE-GWF solution, now a true discontinuity appears along the nodal ($\Gamma$-$M$) direction, from which the quasiparticle weight can be extracted directly (cf. section \ref{sec:spectral_functions}). The transition to the thermodynamic limit is realized by abandoning the real-space cutoff for the correlation functions \textit{and Fourier-transforming all the relevant diagrammatic sums}. As a consequence, the latter can be evaluated for infinite lattice by the Monte-Carlo integration in $\mathbf{k}$-space, which defines the $\mathbf{k}$-DE-GWF method. To end up with a finite-dimensional optimization problem, we, however, retain the cutoff for the range of the effective Hamiltonian parameters $t^\mathrm{eff}_{ij}$ and $\Delta^\mathrm{eff}_{ij}$ (typically up to the three lattice constants). From our experience, due to a rapid decay of the effective hoppings with the relative distance, this is sufficient to accurately reproduce the wave function $|\Psi_0\rangle$ for the non-truncated $\mathcal{H}_\mathrm{eff}$. \textit{These principal alterations define the essence of the $\mathbf{k}$-DE-GWF approach. }

\begin{figure}
  \centering
\includegraphics[width = 0.9\columnwidth]{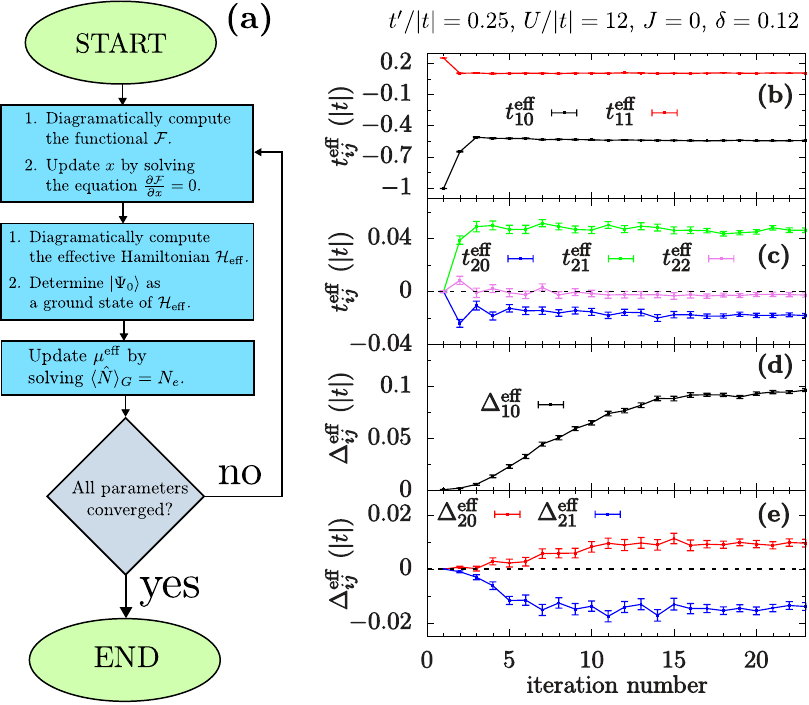}
\caption{(a) The basic flowchart of the $\mathbf{k}$-DE-GWF method. (b)-(e) Selected effective Hamiltonian parameters as a function of the iteration number for the actual self-consistent loop. The model parameters have been taken as $t'/|t| \ 0.25$, $U/|t| = 12$, $J = 0$, and $\delta = 0.12$. The calculations have been performed to the third-order of diagrammatic expansion providing the $d$-wave SC, which results in $\sim 2000$ graphs mapping onto at most $12$-dimensional $\mathbf{k}$-space integrals for single $\mathcal{H}_\mathrm{eff}$ parameter evaluation (cf. \ref{appendix:graphs} for details). Statistical uncertainties are marked by the error bars.}%
\label{fig:flowchart}
\end{figure}

The basic flowchart illustrating the $\mathbf{k}$-DE-GWF method is displayed in figure~\ref{fig:flowchart}(a). Equations (\ref{eq:var_eq_begin})-(\ref{eq:var_eq_end}) are solved in a self-consistent manner, where the correlation parameter $x$, uncorrelated wave function $|\Psi_0\rangle$, and the effective chemical potential $\mu^\mathrm{eff}$ are updated at each step of the  iteration procedure. The technical aspects of the $\mathbf{k}$-space diagrammatic expansion, used to evaluate the functional $\mathcal{F}$ and the effective Hamiltonian $\mathcal{H}_\mathrm{eff}$, are presented in greater detail in \ref{appendix:graphs}. Finally, in figure~\ref{fig:flowchart}(b)-(e) we show the exemplary flow record for a few lowest-range $\mathcal{H}_\mathrm{eff}$ parameters throughout the actual self-consistent loop for the Hubbard model in the $d$-wave SC state to the third order of diagrammatic expansion ($k \leq 3$). The parameters have been set to $t'/|t| = 0.25$, $U/|t| = 12$, $J = 0$, and hole-doping is $\delta = 0.12$. Statistical uncertainties arising from the $\mathbf{k}$-space integration are indicated by the corresponding error bars.

\section{The case of Hubbard model: Discussion of results}
\label{sec:hubbard_model}

Here we apply the $\mathbf{k}$-DE-GWF method to the Hubbard model, obtained by setting $J \equiv 0$ in equation~(\ref{eq:tju-model}) and keeping again only nearest- and next-nearest hopping integrals, $t$ and $t'$, respectively. The variational solution is provided in subsection~\ref{sec:k_de_gwf_solution}, the non-analytic properties of the spectral function moments and their spatial anisotropy are addressed in subsections~\ref{sec:spectral_functions}-\ref{sec:qp_anisotropy}, and the comparison with the available variational Monte-Carlo data is discussed in subsection~\ref{sec:spin_correlations}.

If not stated otherwise, we choose $t'/|t| = 0.25$, appropriate in the context of the high-$T_c$ cuprates. We also set $U/|t| = 12$, which corresponds to the effective antiferromagnetic exchange $J = 4t^2/U = 0.3 |t|$. For $t = -0.3\,\mathrm{eV}$ the Hamiltonian~\ref{eq:tju-model} thus maps onto the $t$-$J$ model with $J \approx 100\,\mathrm{meV}$.

\begin{figure}
  \centering
\includegraphics[width = 0.6\columnwidth]{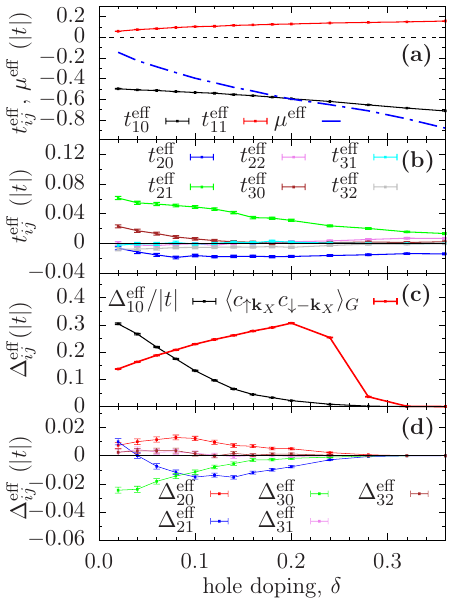} %
\caption{Optimized values of the effective-Hamiltonian parameters (in units of bare $|t|$) as a function of hole doping $\delta$ for the Hubbard model ($t'/t = -0.25$, $U/|t| = 12$), obtained within the $\mathbf{k}$-DE-GWF approach. (a)-(b) Hopping integrals and effective chemical potential $\mu^\mathrm{eff}$. (c)-(d) Effective SC gap components. Additionally, in panel (c) we show anomalous amplitude $\langle c_{\mathbf{k}_X\uparrow} c_{-\mathbf{k}_X\downarrow} \rangle_G \equiv \langle\Psi_G| c_{\mathbf{k}_X\uparrow} c_{-\mathbf{k}_X\downarrow} |\Psi_G\rangle/\langle\Psi_G|\Psi_G\rangle$ at the $X$ point, $\mathbf{k}_X = (\pi, 0)$. The diagrams used by $\mathbf{k}$-DE-GWF algorithm were computed using $2\times10^7$ $\mathbf{k}$-space samples by Cuba Suave algorithm, except for the $\langle c_{\mathbf{k}_X\uparrow} c_{-\mathbf{k}_X\downarrow} \rangle_G$ amplitude, for which sampling at the level of $10^6$ was sufficient. The statistical uncertainties of computed quantities are indicated by the error bars. The red line describes the dome-like behavior, whereas $\Delta^\mathrm{eff}_{10}$ is the amplitude of the leading SC gap component obtained from the effective single-particle Hamiltonian (\ref{eq:heff}).}%
  \label{fig:effective_params_doping}
\end{figure}

\subsection{$\mathbf{k}$-DE-GWF solution}
\label{sec:k_de_gwf_solution}

In figure~\ref{fig:effective_params_doping} we display the $\mathbf{k}$-DE-GWF solution for the parameters of the effective Hamiltonian $\mathcal{H}_\mathrm{eff}$ (cf. equation~(\ref{eq:heff})) as a function of hole doping $\delta \equiv 1 - n$, obtained for the Hubbard model. Panels (a)-(b) detail the effective hopping integrals $t^\mathrm{eff}_{ij}$ and the chemical potential $\mu^\mathrm{eff}$. The reduction of the nearest- and next-nearest hopping integrals $\{t^\mathrm{eff}_{ij}\}$ relative to the their starting values, as seen in panel (a), indicates correlation-induced band narrowing for the projected quasiparticle states. In panel (b) we show few longer-range hopping integrals which arise as an effect of spatially-extended correlations. Note that no hopping beyond the next-nearest neighbors is present in the starting Hamiltonian (\ref{eq:tju-model}). In panels (c)-(d) consecutive amplitudes of the SC-gap components $\{\Delta^\mathrm{eff}_{ij}\}$ are displayed. Their magnitude is smaller by at least an order of magnitude from the leading $\Delta_{10}^\mathrm{eff}$ term.  Additionally, in panel (c) we plot the anomalous $\mathbf{k}$-DE-GWF expectation value $\langle c_{\mathbf{k}_X\uparrow} c_{-\mathbf{k}_X\downarrow} \rangle_G \equiv \langle\Psi_G| c_{\mathbf{k}_X\uparrow} c_{-\mathbf{k}_X\downarrow} |\Psi_G\rangle/\langle\Psi_G|\Psi_G\rangle$ at the $X$ point, $\mathbf{k}_X = (\pi, 0)$, which is a direct measure of the SC correlations contained in the Gutzwiller wave function. This quantity attains its maximal value around $\delta = 0.2$, contrary to the leading effective gap component $\Delta_{10}^\mathrm{eff}$ that monotonously increases as half-filling is approached. Given that the effective gap parameter controls the energy spectrum of projected quasiparticle states (cf. \ref{appendix:effective_hamiltonian}), there is no simple linear scaling between the quasiparicle SC gap $\Delta^\mathrm{eff}_{ij}$ and the Gutzwiller-projected renormalized SC gaps in the resultant correlated state. It is tempting to propose $\Delta^G_{ij} \propto \langle c_{\uparrow i} c_{\downarrow j} \rangle_G$ as a true equilibrium gap parameters and $\Delta^\mathrm{eff}_{ij}$ as a precursor of the pseudogap character. However, detailed analysis of those quantities would require an extension of our results to nonzero temperatures, as well as inclusion of the fluctuating phase factors in $\Delta^\mathrm{eff}_{ij}$. In the outlook (section~\ref{sec:discussion}) we elaborate on this gap duality inherent to the DE-GWF analysis.

\begin{figure}
  \centering
\includegraphics[width = 0.65\columnwidth]{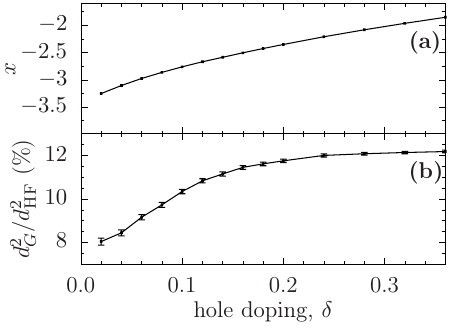} %
\caption{Doping-dependence of (a) correlator-parameter $x$ and (b) probability of the double occupancy $d_G^2 \equiv \langle \hat{n}_{i\uparrow} \hat{n}_{i\downarrow} \rangle_G$, normalized to its Hartree-Fock value $d^2_\mathrm{HF} = n^0_{\uparrow} n^0_{\downarrow}$, for the  Hubbard model with $t'/t = -0.25$, $U/|t| = 12$, obtained within the $\mathbf{k}$-DE-GWF approach. As the half-filling is approached, the parameter changes towards the value $x = -4$ that corresponds to fully projected-out site double occupancies, i.e. the Brinkman-Rice Mott insulating state. Note that within the diagrammatic approach the Brinkman-Rice transition actually never occurs.}%
  \label{fig:var_params_doping}
\end{figure}

In figure~\ref{fig:var_params_doping} we plot the doping evolution of the remaining parameters governing the wave function. Panel (a) exhibits the variational parameter $x$. Close to the half-filling, $x$ steadily decreases and approaches the value $-4$, the limit of the fully projected-out double occupancies, with increasing $U$. Thus, the parameter $x$ cannot be regarded in equations (\ref{eq:t11})-(\ref{eq:i4}) as a perturbation parameter in the ordinary sense and in consequence, the expansions represent an iterative procedure, where the higher-order correlations from the $|\Psi_0\rangle$ state die out steadily with the increasing order $k$ involving the higher-order virtual processes. In panel (b) the probablity of double occupancy $d_G^2$, normalized by its uncorrelated (Hartree-Fock) value $d_\mathrm{HF}^2$, is displayed. A clear correlation-induced $d_G^2$ reduction is observed and is most pronounced for $\delta \rightarrow 0$, again a sign of approaching gradually the quasi-atomic Mott state of electrons. One would require a larger $U$ value to make $d^2_G = 0$ in the $\delta = 0$ limit. However, then the exchange integral $4 t^2 / U$ is too small to match experiment. This contradiction is resolved by introducing the $t$-$J$-$U$ model \cite{SpalekPhysRevB2017_95_024506}.

Having optimized the ground state energy with respect to the variational parameters, we are now in a position to focus on the wave-vector-resolved quantities in the correlated state, as discussed next.

\subsection{Spectral density, Fermi velocities, and the quasiparticle weight}
\label{sec:spectral_functions}

The  consecutive moments of electron spectral function $\mathcal{A}(\mathbf{k}, \omega)$ provide an insight not only into static electronic properties, but also into the low-energy quasiparticle dynamics. Specifically, we consider its first two moments: $\mathcal{M}_0(\mathbf{k}) \equiv \int_{-\infty}^0 \mathcal{A}(\mathbf{k}, \omega) d\omega$ and $\mathcal{M}_1(\mathbf{k}) \equiv \int_{-\infty}^0 \omega \mathcal{A}(\mathbf{k}, \omega) d\omega$. Additionally, we assume the Fermi-liquid character of the spectral function, i.e., $\mathcal{A}(\mathbf{k}, \omega) = Z_\mathbf{k} \delta(\omega - \epsilon^\mathrm{corr}_\mathbf{k}) + \mathcal{A}_\mathrm{inc}(\mathbf{k}, \omega)$, where $Z_\mathbf{k}$ is the so-called quasiparticle weight (inverse of the effective mass renormalization factor),  $\mathcal{A}_\mathrm{inc}(\mathbf{k}, \omega)$ is the incoherent contribution, and $\epsilon^\mathrm{corr}_\mathbf{k}$ denotes the exact quasi-particle dispersion. By substituting this formula into the expressions for the moments, one can show that $\mathcal{M}_0(\mathbf{k})$ exhibits a discontinuity at the Fermi surface of the magnitude $Z_\mathbf{k}$. On the other hand, $\mathcal{M}_1(\mathbf{k})$ is continuous, but develops a cusp at the Fermi surface, where its first derivative jumps by $Z_\mathbf{k} v^\mathrm{corr}_F$ \cite{ParamekantiPhysRevLett2001}. Here $v^\mathrm{corr}_F \equiv \nabla_\mathbf{k}\epsilon^\mathrm{corr}_\mathbf{k} \cdot \hat{n}$, with $\hat{n}$ being a unit vector tangent to the Fermi surface, where $v^\mathrm{corr}_F$ is the value of the Fermi velocity for quasiparticles. The moments of the spectral function can be equivalently expressed by the exact formulas: $\mathcal{M}_0(\mathbf{k}) = n_\mathbf{k}$ and $\mathcal{M}_1(\mathbf{k}) = - \langle c^\dagger_{\mathbf{k}\sigma} [\mathcal{H} - \mu \hat{N}, c_{\mathbf{k}\sigma}] \rangle$ (cf. \ref{appendix:representation_of_m1}). The dynamical properties controlling the low-energy renormalized quasiparticle spectrum are thus encoded in the equal-time commutators with the full Hamiltonian $\mathcal{H}$ which can be evaluated within the $\mathbf{k}$-DE-GWF. The remaining task is to derive the diagrammatic expansion for the first moment $\mathcal{M}_1(\mathbf{k})$ detailed also in \ref{appendix:representation_of_m1}.

\begin{figure}
   \centering
\includegraphics[width = 0.7\columnwidth]{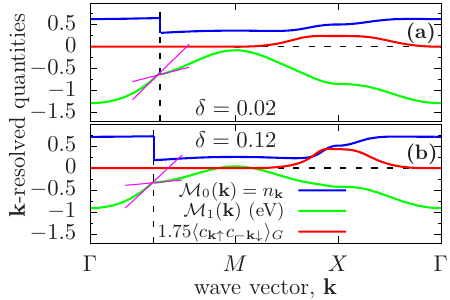} %
\caption{Calculated two leading moments of the spectral function: $\mathcal{M}_0(\mathbf{k}) = n_\mathbf{k}$ (blue lines) and $\mathcal{M}_1(\mathbf{k})$ (green lines)  along the contour $\Gamma$-$M$-$X$-$\Gamma$ for the  Hubbard model ($t'/t = -0.25$ and $U/|t| = 12$) at the hole doping (a) $\delta = 0.02$ and (b) $\delta = 0.12$. The zeroth moment $\mathcal{M}_0(\mathbf{k})$ exhibits discontinuity along the direction $\Gamma$-$M$ equal to the nodal quasiparticle weight. There are no discontinuities along the remaining $M$-$X$-$\Gamma$ part of the contour due to nonzero value of the SC gap (the SC correlations are plotted as the red lines). The first moment exhibits a discontinuity of the first derivative along the $\Gamma$-$M$ line. The linear fits on both sides of this cusp are marked by the magenta lines. For further details see the main text.}%
  \label{fig:moments}
\end{figure}

In figure~\ref{fig:moments} we plot $\mathcal{M}_0(\mathbf{k}) = \ n_\mathbf{k}$ and $\mathcal{M}_1(\mathbf{k})$ along the $\Gamma$-$M$-$X$-$\Gamma$ contour in the Brillouin zone, calculated for the Hubbard model for $t'/t = -0.25$ and $U/|t| = 12$. The top and bottom panels correspond to the hole-doping levels $\delta = 0.02$ and $\delta = 0.12$, respectively. In both cases, the statistical distribution function $n_\mathbf{k}$ (blue line) exhibits a discontinuity along the nodal $\Gamma$-$M$ direction, where the zero-gap quasiparticles are well-defined. The crossing of the Fermi wave vector is indicated by the vertical dashed lines. There is no discontinuity along the $M$-$X$-$\Gamma$ contour, since then the Fermi surface is gapped due to the superconductivity. To further emphasize this point, we illustrate by red lines the anomalous expectation values $\langle c_{\mathbf{k}\uparrow}c_{-\mathbf{k}\downarrow}\rangle_G$ which express the SC correlations in the Gutzwiller wave-function state. It attains the maximal value in the vicinity of the point $X$. The values of the first moment of the spectral function obtained variationally are depicted by the green lines. Note that, along the $\Gamma$-$M$ direction, the slopes of $\mathcal{M}_1(\mathbf{k})$ differ below- and above the Fermi wave vector that is marked by dashed vertical lines; linear fits on both sides are also shown. This is the anticipated feature and physically meaningful discontinuity of the first spectral-function moment.

An important methodological remark is in order at this point. Namely, the variational wave function approach induces certain artifacts in the spectral properties (regardless of the circumstance whether $\mathbf{k}$-DE-GWF, DE-GWF, or VMC method is used to evaluate the expectation values). The reason for this is that the variational function $|\Psi_G\rangle/\sqrt{\langle\Psi_G|\Psi_G\rangle}$ \textit{is not} the exact ground state of Hamiltonian $\mathcal{H}$, and thus inevitably contains an admixture of the excited states. On the other hand, the identity $\mathcal{M}_1(\mathbf{k}) = - \langle c^\dagger_{\mathbf{k}\sigma} [\mathcal{H} - \mu \hat{N}, c_{\mathbf{k}\sigma}] \rangle$ relies on the assumption that the ground state is used to compute the expectation value. Within the trial wave-function approach, the commutator formula for $\mathcal{M}_1(\mathbf{k})$ is thus an approximation that is controlled by an overlap between the variational and the exact ground states. This situation may lead to two types of artifacts in the first variational moment: \textit{(i)} small discontinuities of $\mathcal{M}_1(\mathbf{k})$ at the Fermi surface that can be seen by a careful inspection of Figs.~\ref{fig:moments}(a)-(b). Note that, due to limited $\mathbf{k}$-space resolution, VMC does not prove conclusively that there is a jump of $\mathcal{M}_1(\mathbf{k})$ at the Fermi surface \cite{ParamekantiPhysRevLett2001}; \textit{(ii)} $\mathcal{M}_1(\mathbf{k})$ might become positive around some points in the Brillouin zone (cf. small positive values of $\mathcal{M}_1(\mathbf{k})$ near  $M$ point in figure~\ref{fig:moments}(b)). From this perspective, the zeroth moment $\mathcal{M}_0(\mathbf{k})$ is different, as it equals to the statistical distribution function $n_\mathbf{k}$, which guarantees fulfillment of several exact relations by construction, e.g., $0 \leq \mathcal{M}_0(\mathbf{k}) \leq 1$. This remains true even if the trial function is not the exact ground state. In \ref{appendix:artifacts} we consider the exactly soluble non-interacting limit, and show explicitly how the use of approximate ground states generates those artifacts. In such a situation, we also demonstrate that the physically meaningful jump of the first derivative of $\mathcal{M}_1(\mathbf{k})$ at the Fermi surface is weakly affected by altering the wave function. Finally, we argue that, for generic case, the discontinuity of $\mathcal{M}_1(\mathbf{k})$ might lead to the circumstance that variationally computed $v^\mathrm{corr}_F$ reflects the slope of quasiparicle dispersion away from Fermi surface (as we discuss next; this has implications for the detailed analysis of the photoemission data).

\subsection{Comparison with the experimental results and their interpretation}
\label{sec:comparison_experiment}

\begin{figure}
   \centering
\includegraphics[width = 0.6\columnwidth]{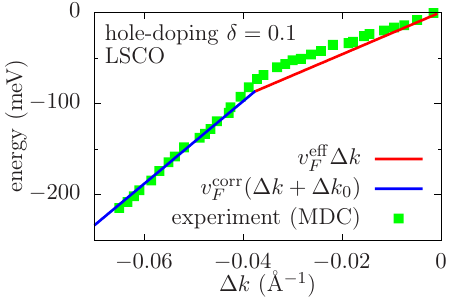} %
\caption{Kink in the ARPES spectra for LSCO as obtained from the momentum distribution curves (MDCs) at hole-doping level $\delta = 0.1$, based on the data of \cite{ZhouNature2003} (green squares). Here $\Delta k$ denotes wave vector relative to the Fermi surface along the nodal $\Gamma$-$M$ direction in the Brillouin zone. The red line represents the dispersion $\epsilon_k^\mathrm{eff} = v_F^\mathrm{eff} \Delta k$ with $v_F^\mathrm{eff}$ extracted from the effective Hamiltonian $\mathcal{H}_\mathrm{eff}$ that arises within the $\mathbf{k}$-DE-GWF (calculations have been performed for the Hubbard model with $t'/t = -0.25$, $U/|t| = 12$, and $\delta = 0.1$). The blue line shows the slope related to the ``correlated'' velocity $v^\mathrm{corr}_F$ that is obtained from the spectral-function moments, also within the $k$-DE-GWF approach. Specifically $\epsilon_k = v^\mathrm{corr}_F(\Delta k + \Delta k_0)$ with one fitting parameter $\Delta k_0 \approx 0.0184\,$\AA${}^{-1}$ (cf. the discussion in the text).}
  \label{fig:kink}
\end{figure}

\begin{figure}
   \centering
\includegraphics[width = 0.6\columnwidth]{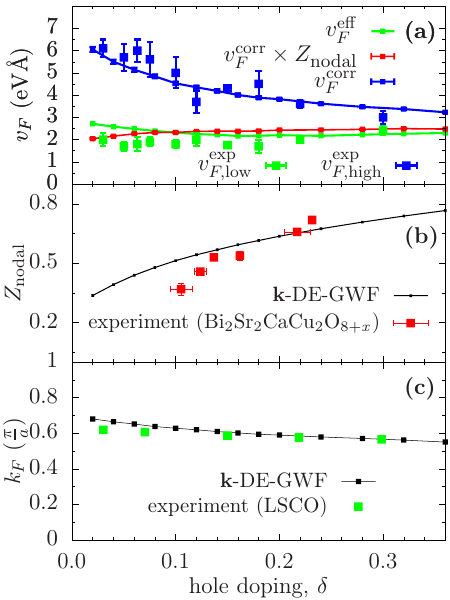} %
\caption{Nodal quasiparticle properties for the  Hubbard model with $t'/t = -0.25$ and $U/|t| = 12$, plotted as a function of hole doping $\delta$ and calculated within the $\mathbf{k}$-DE-GWF approach: (a) Effective (green line) and correlated (blue line) Fermi velocities $v_F^\mathrm{eff}$ and $v_F^\mathrm{corr}$, obtained from the energy spectrum of the effective Hamiltonian $\mathcal{H}_\mathrm{eff}$ and $\mathcal{H}$, averaged with $|\Psi_0\rangle$ and $|\Psi_G\rangle$, respectively, providing the corresponding expressions for the two leading moments of the spectral function. These two velocities scale differently with $\delta$ and are approximately connected by the relation $v_F^\mathrm{eff} = Z_\mathrm{nodal} v_F^\mathrm{corr}$ (cf. the red line in panel (a) which depicts $Z_\mathrm{nodal} \times v_F^\mathrm{corr}$). The blue- and green squares show the Fermi velocity obtained from the slope of experimental \cite{ZhouNature2003,MatsuyamaPhysRevB2017} low-energy and high-energy photoemission spectra for LSCO ($v_{F, \mathrm{low}}^\mathrm{exp}$ and $v_{F, \mathrm{high}}^\mathrm{exp}$, respectively). In panel (b) we display the calculated (black line) and experimental \cite{JohnsonPhysRevLett2001,RanderiaPhysRevB2004} (for BSCCO) evolution of the nodal quasiparticle weight. In panel (c) the calculated (black line) and experimental \cite{HashimotoPhysRevB2008} (for LSCO) Fermi wave-vector values are displayed.}
  \label{fig:velocities}
\end{figure}

Keeping in mind the approximate nature of the wave function, we extract the velocity $v^\mathrm{corr}_F$ characterizing the correlated state from the discontinuity of the first derivative of $\mathcal{M}_1(\mathbf{k})$ and the jump of $\mathcal{M}_0(\mathbf{k})$. We also point out that the additional, effective Fermi velocity, $v_F^\mathrm{eff}$, can be defined using the single-particle energy spectrum $\epsilon_\mathbf{k}^\mathrm{eff}$ extracted from the effective Hamiltonian $\mathcal{H}_\mathrm{eff}$ (i.e., from $\mathcal{H}_\mathrm{eff} |\Psi_0\rangle = \epsilon_\mathbf{k}^\mathrm{eff} |\Psi_0\rangle$). Such a dichotomy is not unphysical as two velocity scales are indeed observed in the photoemission spectra of high-$T_c$ cuprates as ubiquitous kinks. Namely, experimental $v_{F, \mathrm{low}}^\mathrm{exp}$ and $v_{F, \mathrm{high}}^\mathrm{exp}$ reflect the slope of the dispersion curve above and below the kink, respectively. This is illustrated in figure~\ref{fig:kink}, where we compare the data of reference~\cite{ZhouNature2003} for LSCO with the $\mathbf{k}$-DE-GWF solution at fixed doping $\delta = 0.1$. Here we have specified the values of the nearest-neighbor hopping $t = -0.35\,\mathrm{eV}$ and the lattice spacing $a = 3.78\,\mbox{\AA}$. The considered Hubbard Hamiltonian with $U/|t| = 12$ maps then onto the $t$-$J$ model with $J \approx 117\,\mathrm{meV}$, that is in the experimental range for the high-$T_c$ cuprates. Remarkably, the relative wave vector $\Delta k \equiv k - k_F$ needs to be shifted by $\Delta k_0 \approx 0.0184\,$\AA${}^{-1}$ for the higher-energy solution to collapse onto the experimental points. We attribute this circumstance to the non-zero admixture of excited states in the Gutzwiller wave function that can be also directly related to the discontinuity of the first spectral moment $\mathcal{M}_1(\mathbf{k})$. By making use of the estimate, derived in \ref{appendix:artifacts}, we would get the wave-vector mismatch $\sim \Delta \mathcal{M}_1 / (Z v_F^\mathrm{corr}) \approx 0.013\,$\AA${}^{-1}$ which is not far from $0.0184\,$\AA${}^{-1}$. Physically, we can write for the two excitations branches along the nodal direction that $\epsilon_\mathbf{k}^\mathrm{eff} \approx v^\mathrm{eff}_F (k - k_F)$ for the quasiparticle branch and $\epsilon_\mathbf{k}^\mathrm{corr} \approx v^\mathrm{corr}_F (k - k_F^\mathrm{corr})$ or, alternatively, $\epsilon_\mathbf{k}^\mathrm{corr} \approx v_F^\mathrm{corr} (k - k_F) + v_F^\mathrm{corr} \Delta k_0$, where $\Delta k_0 \equiv k_F - k_F^\mathrm{corr}$. Effectively, the higher excitation branch is shifted with respect to the quasiparticle one by $\Delta k_0$.

In figure~\ref{fig:velocities} we present the most important result of the paper, namely the hole doping dependence of the calculated $v_F^\mathrm{corr}$ and other quasi-particle properties (once again, for $t = -0.35\,\mathrm{eV}$ and $a = 3.78\,\mbox{\AA}$).  In panel (a) two distinct quasiparticle velocity scales are shown: \textit{(i)} Effective velocity $v_F^\mathrm{eff}$ (green line). As is detailed in \ref{appendix:effective_hamiltonian}, $v_F^\mathrm{eff}$ coincides with the Fermi velocity of projected quasiparticles. \textit{(ii)} The so-called correlated velocity $v_F^\mathrm{corr}$ (blue line), extracted from the singular part of the second spectral-function moments, as described above. The basic distinction between those quantities  is that the latter does not rely on the validity of the projected quasiparticle ansatz for the excited states. Note that $v_F^\mathrm{eff}$ and $v_F^\mathrm{corr}$ scale differently with the doping and are approximately connected by the relation $v_F^\mathrm{eff} \approx Z_\mathrm{nodal} \times v_F^\mathrm{corr}$, where $Z_\mathrm{nodal}$ denotes the nodal quasiparticle weight. The last feature of the results is illustrated in figure~\ref{fig:velocities}(a) by the red line. The full squares show experimental velocities $v_{F, \mathrm{low}}^\mathrm{exp}$ and $v_{F, \mathrm{high}}^\mathrm{exp}$, obtained from ARPES momentum distribution curves (MDCs) \cite{ZhouNature2003,MatsuyamaPhysRevB2017}. The $\mathbf{k}$-DE-GWF result matches quantitatively the experimental values in entire doping range. Even though within the present approach we are unable to explicitly obtain a systematic evolution of the quasiparticle properties as a function of energy, this agreement indicates that the low-energy excited states can be described approximately by the projected BCS wave function, whereas at higher energies the overlap between the exact- and projected-quasi-particle states systematically decreases. The applicability of our projected quasiparticle ansatz for the description of low-lying excitations is independently supported by Lanczos \cite{OhtaPhysRevLett1994} and VMC \cite{YunokiPhysRevB2006} studies that reproduce sharp features of the anomalous spectral function in the $t$-$J$ model, in agreement with our renormalized  BCS theory. However, this last approach does not yield pronounced kinks in the dispersion spectra obtained in the photoemission. The correlation-driven bending of the dispersion curve at a threshold energy can be argued within the recently proposed concept of \textit{extremely correlated Fermi liquid} (ECFL) \cite{ShastryPhysRevLett2011,ShastryPhysRevB2013,MatsuyamaPhysRevB2017} which has been based on a perturbative treatment of the double-occupancy projection combined with equations-of-motion method for the Matsubara Green's functions. Note that neither $\mathbf{k}$-DE-GWF nor ECFL relies on the presence of any long-wavelength bosonic modes to generate two well-defined  velocity scales. Instead, they  result from local correlations. This is also independently supported by the dynamical mean-field theory calculations \cite{ByczukNatPhys2007}. To complete the analysis of the quasiparticle properties, we plot in figure~\ref{fig:velocities}(b) the calculated nodal quasiparticle weight $Z_\mathrm{nodal}$ (black line). The red squares are experimental points for BSCCO, for which the values of $Z_\mathrm{nodal}$ are available \cite{JohnsonPhysRevLett2001,RanderiaPhysRevB2004}. Panel (c) shows the obtained doping-evolution of the Fermi wave vector (black line). The latter depends weakly on the hole concentration, also in agreement with experiments \cite{HashimotoPhysRevB2008} on LSCO (solid squares).

\subsection{A brief methodological remark}
\label{sec:methodological_remark}

One should mention that the data for $v_{F, \mathrm{low}}^\mathrm{exp}$ and $k_F$ (cf. figure \ref{fig:velocities}a and c) have been previously rationalized qualitatively within the $t$-$J$-$U$ model (cf. figures 6 and 7a in \cite{SpalekPhysRevB2017_95_024506}) as coming from the excitations of the effective single-particle Hamiltonian $\mathcal{H}_\mathrm{eff}$. There, we have taken $J/|t| = 0.25$, but $U/|t| \approx 22$, much larger than that used here, which is $U/|t| = 12$. However, the kinetic exchange in \cite{SpalekPhysRevB2017_95_024506} was included explicitly there and therefore, the principal role of the Hubbard term was to introduce strong correlations (suppress the double occupancies). Hence, the differences can be understood easily. Namely, the two formulations: the Hubbard and $t$-$J$-$U$ models (the latter being the large-$U$ limit of either the Hubbard \cite{ChaoJPhysC1977} or extended Hubbard forms \cite{SpalekPhysStatSolidiB1981}) represent slightly different ways of mapping the multi-band model onto the one-band effective model. It is gratifying that the two models provide the same principal physics. Nevertheless, it should be noted that the DE-GWF solution for the Hubbard model does not provide the proper reduction of the kinetic energy in the SC phase and in the underdoped regime, but the $t$-$J$-$U$ does \cite{SpalekPhysRevB2017_95_024506}. The kinetic-energy reduction seems to present itself as one of the most stringent tests of various methods and models, when combined with simultaneous quantitative analysis (i.e., for the fixed parameters, $|t|$, $J$, $U$, etc.) of other experimental quantities \cite{SpalekPhysRevB2017_95_024506}. The other is the emergence of the pseudogap as driven purely by correlations (cf. also \cite{BrangancaPhysRevLett2018}). It is tempting to associate the pseudogap with our parameter $\Delta^\mathrm{eff}_\mathbf{k}$ (see below).

\subsection{Anisotropy of spectral quasiparticle properties}
\label{sec:qp_anisotropy}

\begin{figure}
  \centering 
\includegraphics[width = 0.6\columnwidth]{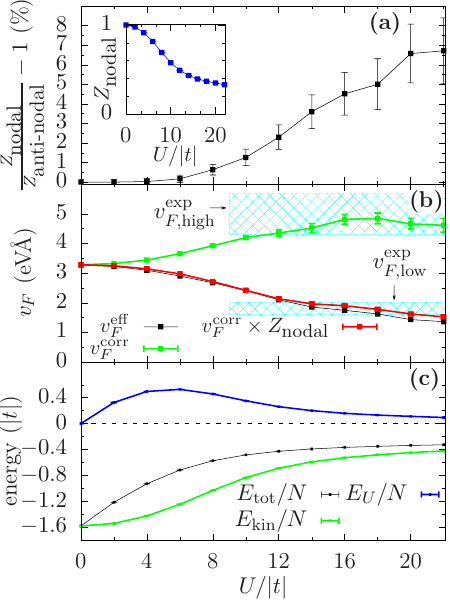} %
\caption{Selected normal-state properties of the Hubbard model for $t'/|t| = 0.25$ and $\delta = 0.1$, as a function of the on-site repulsion $U$. (a) Directional anisotropy between spectral weights of the  nodal ($Z_\mathrm{nodal}$) and anti-nodal quasiparticles near the $X$ point ($Z_\mathrm{anti\mbox{-}nodal}$). The inset shows the value of $Z_\mathrm{nodal}$. (b) Calculated effective and correlated Fermi-velocity $v_F^\mathrm{eff}$ and $v_F^\mathrm{corr}$ scales. The shaded regions correspond to the experimental velocities $v_F^\mathrm{low} = 1.8 \pm 0.2 \,\mathrm{eV\mbox{\AA}}$ and $v_F^\mathrm{high} = 5.0 \pm 0.7\, \mathrm{eV\mbox{\AA}}$, determined from low- and high-energy photoemission spectra for LSCO from ARPES momentum distribution curves (MDCs) \cite{ZhouNature2003,MatsuyamaPhysRevB2017}. Their height reflects uncertainty of the measured values. (c) Calculated kinetic- and Hubbard-interaction energies (blue and green lines, respectively). Black line shows the total energy per site. The non-monotonic behavior of the potential energy marks a crossover between weak- and strong-correlation regimes.}
  \label{fig:variable_u}
\end{figure}

So far we have focused on the nodal quasiparticles that are gapless in the $d$-wave SC state. We now turn to the discussion of the normal state, which allows to address directional dependence of the spectral quasiparticle properties. This is essential feature, since in the $d$-wave SC state, considered above, the distribution function $n_\mathbf{k}$ exhibits a Fermi ridge only along the nodal lines. We fix the hole-doping level at $\delta = 0.1$ and monitor how the spectral anisotropy emerges as a function of the on-site repulsion $U$.

In figure~\ref{fig:variable_u}(a) we plot the relative spectral weight anisotropy parameter $Z_\mathrm{nodal}/Z_\mathrm{anti\mbox{-}nodal} - 1$ with respect to the nodal quasiparticles on the $\Gamma$-$M$ line and those located on the $M$-$X$-$\Gamma$ contour, close to the $X$ point (we call the latter anti-nodal quasiparticles).  Both $Z_\mathrm{nodal}$ and $Z_\mathrm{anti\mbox{-}nodal}$ have been extracted from the statistical distribution functions $n_\mathbf{k}$. For $U = 0$, these two spectral weights are equal to unity and are direction independent. As the interactions are turned on, a small but definite anisotropy develops between $Z_\mathrm{nodal}$ and $Z_\mathrm{anti\mbox{-}nodal}$; the spectral weight concentrates along the nodal direction. Note that a small difference $< 10\%$ remains well within the resolution of the $\mathbf{k}$-DE-GWF approach (statistical uncertainties are indicated on the plot), contrary to the real-space result, where the Gibbs-type oscillations occur on even larger scale (cf. figure~\ref{fig:comparison_mz}). Comparable magnitudes of the anisotropy have been reported for the $t$-$J$ model for the Gutzwiller-projected wave function \cite{BieriPhysRevB2007}. Inset in panel (a) shows the value of $Z_\mathrm{nodal}$ that is rapidly suppressed with the increasing $U$, indicating the loss of quasiparticle-state coherence. 

In figure~\ref{fig:variable_u}(b) we present analysis complementary to that of figure~\ref{fig:velocities}(a), i.e., the interaction-dependence of the effective and correlated velocities, $v_F^\mathrm{eff}$ and $v_F^\mathrm{corr}$, respectively. In the non-interacting ($U$ = 0) case, they are equal, $v_F^\mathrm{eff} = v_F^\mathrm{corr}$. As the interactions increase, these two velocities split and, around $U/|t| \sim 12$, fall into the corresponding experimental ranges for the values below and above the kink in the dispersion relation (shaded areas). Finally, in panel (c) we plot the contributions to the total ground-state energy per site from the Hubbard ($E_U$) and kinetic ($E_\mathrm{kin}$) terms. The sum of the two, $E_\mathrm{tot}$, is depicted as a black line. The crossover from the weak- to strong-correlation regime is reflected by a non-monotonic interaction-dependence of $E_U$. For small $U$ the original Fermi-sea ground state is robust to interactions and the potential energy increases roughly as $U n^0_\uparrow n^0_\downarrow$. For sufficiently large $U$, however, it is energetically favorable to qualitatively reorganize the state, as is reflected in the trial wave function via essential reduction of the local double occupancies. This leads to a reduction of $E_U$, but at the same time, to a partial reduction of the negative kinetic contribution $E_\mathrm{kin}$. Note that the state reorganization takes place around $U \sim W \approx 8|t|$ which defined the crossover from moderately to strongly correlated regime.

\subsection{Instantaneous spin correlations: comparison with VMC}
\label{sec:spin_correlations}

\begin{figure}
  \centering
\includegraphics[width = 1.0\textwidth]{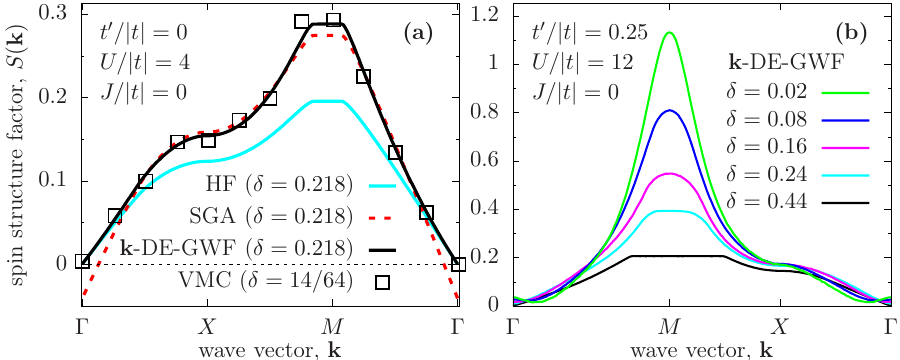} %
\caption{(a) Instantaneous (equal-time) spin structure factor $S(\mathbf{k})$ along the $\Gamma$-$X$-$M$-$\Gamma$ contour in the Brillouin zone for the Hubbard model ($t'/|t| = 0$, $U/|t| = 4$, $J/|t| = 0$, and $\delta = 0.218$). The $\mathbf{k}$-DE-GWF and the (non-diagrammatic) Gutzwiller-approximation results are depicted by solid black and dashed red lines, respectively.  Note the unphysical behavior of the latter near the $\Gamma$ point. The solid blue line represents the Hartree-Fock result. Additionally, the variational Monte-Carlo (VMC) data of reference~\cite{OtsukaJPhysSocJapan1992} are displayed by squares (the VMC calculations have been actually performed for $14$ electrons in $8 \times 8$ lattice that yields $\delta = 0.21875$). (b) Doping-dependence of $S(\mathbf{k})$ across the Brillouin zone for $t'/t = 0.25$, $U/t = 12$, and $J/|t| = 0$ in the $d$-wave SC phase. The $\mathbf{k}$-DE-GWF calculations have been performed to the third order of the diagrammatic expansion ($k \leq 3$).}%
  \label{fig:structure_factor}
\end{figure}

Within the $\mathbf{k}$-DE-GWF approach we can determine a variety of instantaneous (equal-time) correlations, e.g., spin, charge, or nematic. Here we restrict to the spin structure factor $S(\mathbf{k}) \equiv \langle \hat{S}_\mathbf{k}^z \hat{S}_{-\mathbf{k}}^z \rangle_G = \frac{1}{4} \langle \hat{S}_\mathbf{k}^{+} \hat{S}_{-\mathbf{k}}^{-} + \hat{S}_\mathbf{k}^{-} \hat{S}_{-\mathbf{k}}^{+}  \rangle_G$, which is also available from extensive variational Monte-Carlo data and thus can be compared to those obtained within our mehtod. In figure~\ref{fig:structure_factor}(a) the instantaneous spin structure factor for the Hubbard model with $t'/|t| = 0$, $U/|t| = 4$, $J/|t| = 0$, and $\delta = 0.218$ is calculated to the third expansion order in the normal state and depicted by the black solid line. This result agrees well with the variational Monte-Carlo data of reference~\cite{OtsukaJPhysSocJapan1992} (open squares). We emphasize that VMC provides a discrete set of points in $\mathbf{k}$-space as a consequence of finite lattice considered ($8 \times 8$ sites in this case), whereas the $\mathbf{k}$-DE-GWF yields a continuous curve, since it works in the thermodynamic limit. The red dashed line shows the statistically-consistent Gutzwiller approximation (SGA) result, which is obtained by discarding all non-local diagrammatic contributions or, equivalently, by taking the limit of infinite number of spatial dimensions (cf. reference~\cite{KaczmarczykNewJPhys2014} for additional discussion of the relation between SGA and DE-GWF). The unphysical behavior of the SGA results is apparent near the $\Gamma$ point, where $S(\mathbf{k})$ becomes negative, contrary to the full $\mathbf{k}$-DE-GWF solution, where no such an artifact appears. This result underlines the necessity of including higher-order diagrammatic contributions to reliably describe $\mathbf{k}$-resolved structure factors. The blue solid line is the Hartree-Fock result. By comparing the latter with either $\mathbf{k}$-DE-GWF or VMC one can see that the spin correlations are substantially enhanced by electronic correlations. Note that $S(\mathbf{k})$ is peaked around the $M$ point, indicating considerable antiferromagnetic correlations in the Gutzwiller wave function. We can ascribe this enhancement to the inclusion of local spin fluctuations. Note, however, that the long-wavelength paramagon-type excitations are still not included. The latter are expected to play an important role in the direct vicinity of half-filling, where the antiferromagnetic and magnetic inhomogeneous phases (not considered here) are approached.

In figure~\ref{fig:structure_factor}(b) we display $S(\mathbf{k})$ for the Hubbard model with the same  parameters as those in previous sections ($t'/|t| = 0.25$, $U/|t| = 12$, $J/|t| = 0$), calculated within the $\mathbf{k}$-DE-GWF approach in the $d$-wave SC state up to the third order of diagrammatic expansion. The antiferromagnetic fluctuations near the $M$ point undergo approximately $6$-fold enhancement as the doping level decreases from $\delta = 0.44$ to $\delta = 0.02$, signaling the tendency to antiferromagnetic instability. Remarkably, due to emergence of the pseudogap in the projected quasiparticle spectra (evidenced by the monotonous increase of the effective gap $\Delta_{10}^\mathrm{eff}$ towards half-filling, seen in figure~\ref{fig:effective_params_doping}(c)), the $S(\mathbf{k})$ becomes increasingly smooth as the $\delta = 0$ limit is approached. Specifically, the two cusps around the $M$ point, visible for $\delta = 0.44$, are absent for $\delta = 0.02$.  

\section{Outlook and a further interpretation of results}
\label{sec:discussion}

In this paper we have introduced the $\mathbf{k}$-DE-GWF variational method for the systems of correlated and itinerant fermions that is based on a systematic treatment of the Gutzwiller partial-projection operators directly in the thermodynamic limit. Its main advantage, as compared to other variational techniques, such as the variational Monte-Carlo (VMC) or DE-GWF, is that it is free of the finite-range diagram summation artifacts which are smearing out the singular character of wave-vector-resolved quantities, such as jump-discontinuities and cusps of the consecutive spectral-function moments. The universal feature of the method is that we start from the real-space description of the interparticle correlations, but evaluate systematically the relevant averages directly in the reciprocal ($\mathbf{k}$) space. We have focused on those singular features and used
them to extract quasiparticle energies and, related to them, Fermi velocity scales $v^\mathrm{corr}_F$ and $v_F^\mathrm{eff}$ appearing in a natural manner as features of $|\Psi_G\rangle$ and $|\Psi_0\rangle$, respectively. On the other hand, the remaining $\mathbf{k}$-independent characteristics ($\Delta^\mathrm{eff}_{ij}$, $E_G$, etc.) are close in value when calculated with either DE-GWF or $\mathbf{k}$-DE-GWF (cf.~\ref{appendix:real_space_scaling}).

Thus, the present approach allows us to improve the description of system properties as a function of the hole doping, obtained within the real-space diagrammatic expansion (DE-GWF), by eliminating the spatial cutoff that limits the range of interelectronic correlations taken into account to few coordination spheres. The technical change of going directly to the reciprocal-space evaluation of the consecutive expansion coefficients proved to be essential in the following aspects: \textit{(i)} the discontinuities of the distribution function $n_\mathbf{k}$ are now properly defined and determined, and \textit{(ii)} the correlated state characteristics such as $\Delta^G$ is shown to be characterizing the fully correlated ground-state. However, $v_F^\mathrm{corr}$ describes the high-energy excited state and thus requires more energy than those very close to the Fermi surface. The quasiparicle Hamiltonian $\mathcal{H}_\mathrm{eff}$ and $|\Psi_0\rangle$ represents the low-energy \textit{BCS-type theory}, with the $d$-wave form of the non-renormalized order parameter $\Delta^\mathrm{eff}$, whereas the state $|\Psi_G\rangle$ provides intrinsic characteristics of the fully correlated state. The basic question is whether those two states describe experimental dynamical properties \textit{on two different energy scales} \cite{KondoPhysRevLett2007,ChatterjeePNAS2011,KondoNature2009}, the pseudogap ($\Delta^\mathrm{eff}$), Fermi velocity ($v^\mathrm{eff}$), and the correlated $\Delta^G$ and $v_F^\mathrm{corr}$, respectively. 

\begin{figure}
  \centering
\includegraphics[width = 0.7\textwidth]{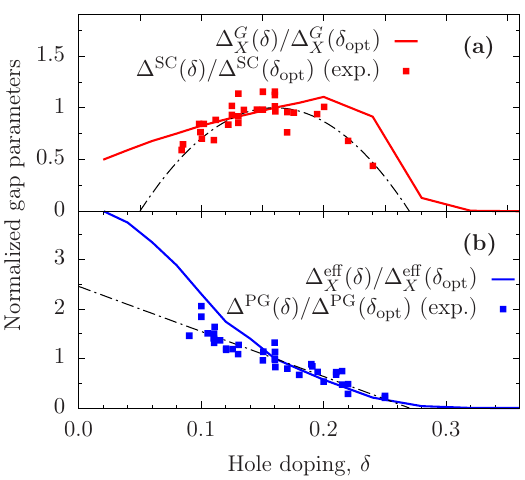} %
\caption{Two distinct gap parameters at the $X$-point ($\mathbf{k} = (\pi, 0)$) as a function of hole concentration $\delta$, both normalized to their values at optimal doping $\delta_\mathrm{opt} \approx 0.16$. Solid colored lines represent the $\mathbf{k}$-DE-GWF result for the correlated gap $\Delta^G_\mathbf{k} \propto \langle c_{\uparrow \mathbf{k}} c_{\downarrow -\mathbf{k}}\rangle_G$ (panel (a)) and effective wave-vector-resolved gap $\Delta^\mathrm{eff}_\mathbf{k}$ (panel (b)). The squares are experimental data for the  the SC gap ($\Delta^{SC}$) and pseudogap ($\Delta^\mathrm{PG}$) obtained by various experimental probes, taken from reference~\cite{HufnerRepProgPhys2008}. The dot-dashed lines represent the guide for the eye of the data trend. The parameters are the same as those taken for the computation of Fermi velocities. The gap $\Delta^G_\mathbf{k}$ vanishes for large values of $U$ in the $\delta \rightarrow 0$ limit \cite{KaczmarczykNewJPhys2014}.}%
  \label{fig:gap_pseudogap}
\end{figure}

Although our main purpose in this paper was to discuss dynamic quantities, we have attempted also to compare the results presented in figure~\ref{fig:effective_params_doping}(c) for $\Delta^G_\mathbf{k} \propto \langle c_{\uparrow \mathbf{k}} c_{\downarrow \mathbf{-k}} \rangle_G$ and $\Delta^\mathrm{eff}_\mathbf{k}$ with the experimental data accumulated in \cite{HufnerRepProgPhys2008} and concerning the SC gap ($\Delta^\mathrm{SC}$) and pseudogap ($\Delta^\mathrm{PG}$), respectively. In figure~\ref{fig:gap_pseudogap} we plot those gaps, normalized to their values for the optimal doping, against the data points for $\Delta^\mathrm{SC}$ -- (a) and $\Delta^\mathrm{PG}$ -- (b), both as a function of hole-doping.  The overall data trends versus $\delta$ are well reproduced, except the doping dependence $\Delta^G_\mathbf{k}$ in the regime $\delta \leq 0.1$, since the value of the Hubbard-$U$ takes as $U = 12|t|$ is to small to suppress $\Delta^G_\mathbf{k}(\delta)$ when approaching the Mott insulator limit. A better quantitative parametrization of the gap is obtained when one goes beyond the analyzed here in detail Hubbard model, e.g., by using the $t$-$J$-$U$ model in the same scheme \cite{SpalekPhysRevB2017_95_024506}. Similar scenario and an alternative rationalization for the $t$-$J$-$U$-model extensions is provided by the recently proposed \cite{WysokinskiPhysRevB2017,WysokinskiPhysRevB2017_2} adaptive canonical transformation as applied to the pure Hubbard Hamiltonian. A detailed analysis of such an involved model with the $\mathbf{k}$-DE-GWF is beyond our present computational capabilities. However, we point out that recent studies of the gap \cite{YoshidaJPhysSocJapan2012,HashimotoNatPhys2014}, based on extrapolation from the near-nodal direction, suggest that it flattens out in the underdoped region, in contrast to dome-like trend seen in figure~\ref{fig:gap_pseudogap}(a). The latter could yield a better agreement with our calculation, though a separate analysis would be necessary to confirm this. Note also that, with the help of the comparison in figure~\ref{fig:gap_pseudogap}(b), it is tempting to interpret $\Delta^\mathrm{eff}_\mathbf{k}(\delta)$ as that emulating the  pseudogap. However, this interpretation attempt should be tested much more accurately, as for example, one would have to attach random phase fluctuations to $\Delta^\mathrm{eff}_{ij}$ and thus test whether this quantity can play the role of pseudogap without spontaneous breaking of the symmetry leading to the fully developed SC state. For the time being, it is merely a curious observation.

The above dichotomy between correlated and effective gaps, $\Delta^G_{ij}$ an $\Delta^\mathrm{eff}_{ij}$, as well as the emergence of two distinct velocity scales, $v_F^\mathrm{corr}$ and $v_F^\mathrm{eff}$, seems to be inherent to the employed diagrammatic variational method going beyond the mean-filed (SGA) approximation. However, due to the lack of the full dynamics in this approach, it does not provide the crossover between corresponding energy regimes in a systematic manner. This limitation can be circumvented by using the Green's function techniques that allow for treating electronic self-energy in a self-consistent manner \cite{AvellaPhysRevB2007,PlakidaPhysicaC2016,KorshunovEurPhysJB2007,OvchinnikovBook2004}.  The two mentioned approaches can be, to some extent, treated as complementary: Green's function methods are well suited for studying dynamical effects, but usually one needs to employ more approximate treatment of electronic correlations. Inverse is true for the DE-GWF and $\mathbf{k}$-DE-GWF techniques. Namely, the correlations, static statitstical properties, and the dispersion relation for fermionic excitations are well reproduced, but the detailed dynamics (i.e., the excitation lifetime) is absent. Combining the advantages of both schemes would provide a decisive progress.

\section*{Acknowledgments}

We thank Adam Kami{\'n}ski for the discussion of experimental photoemission-related aspects of the superconducting gap and pseudogap, and Maxim M. Korshunov for turning our attention to detailed features of the Hubbard operator Green-function techniques. This work was supported by MAESTRO Grant No. DEC-2012/04/A/ST3/00342 from Narodowe Centrum Nauki (NCN).

\appendix

\section{Interpretation of the effective Hamiltonian}
\label{appendix:effective_hamiltonian}

In this Appendix we provide a rigorous physical interpretation of the effective Hamiltonian $\mathcal{H}_\mathrm{eff}$. Specifically, we show that, up to well controlled corrections $o(x^2)$, the quasiparticle spectrum of $\mathcal{H}_\mathrm{eff}$ is given by expectation values $\epsilon_\mathbf{k\sigma}^\mathrm{eff} =  \langle \Psi_{G\mathbf{k}\sigma}| (\mathcal{H} - \mu\hat{N}) |\Psi_{G\mathbf{k}\sigma}\rangle - \Omega_G$, where $|\Psi_{G\mathbf{k}\sigma}\rangle \equiv P_G \gamma^\dagger_{\mathbf{k}\sigma} | \Psi_\mathbf{k}\rangle \times || P_G \gamma^\dagger_{\mathbf{k}\sigma} |\Psi_0\rangle||^{-1}$ and $\gamma^\dagger_{\mathbf{k}\sigma}$ creates a Bogoliubov quasiparticle. The norm $|| P_G \gamma^\dagger_{\mathbf{k}\sigma} |\Psi_0\rangle ||$ ensures proper normalization. Previously, the validity of this statement was demonstrated for the limit of infinite number of spatial dimensions that maps onto truncated form of zeroth-order DE-GWF/$\mathbf{k}$-DE-GWF approximation (cf. Appendix of reference~\cite{BunemannPhysRevB2003}). By using $\mathcal{H}_\mathrm{eff}$ to evaluate physical quantities, such as nodal Fermi velocity or free energy at finite temperatures one thus implicitly relies on the projected quasi-particle ansatz for excited states. This should be contrasted with the analysis based on the moments of the electron spectral function that is not biased in this manner. 

\subsection{Notation}

Let us first define two expectation values $N \equiv \langle \Psi_0 | P_G  (\mathcal{H} - \mu \hat{N}) P_G | \Psi_0\rangle$ and $D \equiv \langle \Psi_0 | P_G^2 | \Psi_0\rangle$. The  grand potential is thus written as $\Omega_G = N/D$ and the variational energy reads $E_G = N/D + \mu N_e$. If $N$ and $D$ are evaluated by means of Wick's theorem, the potential $\Omega_G$ becomes a function of uncorrelated density matrix elements $P_l = \langle \Psi_0| \hat{o}_1 \hat{o}_2|\Psi_0\rangle$, where $\hat{o}_1$ and $\hat{o}_2$ are fermionic creation or annihilation operators on the lattice, i.e., $\Omega_G = \Omega_G(\{P_l\})$. The index $l$ runs over all possible operator pairs. In this Appendix, contrary to the main text, we use a uniform notation for the normal ($\langle\Psi_0|c^\dagger_{i\sigma} c_{j\sigma} |\Psi_0\rangle$) and anomalous ($\langle\Psi_0|c_{i\uparrow} c_{j\downarrow} |\Psi_0\rangle$) expectation values, both of which are special cases of general expression $\langle\Psi_0|\hat{o}_1 \hat{o}_2 |\Psi_0\rangle$. This makes the formulas appearing in the reasoning more compact. To facilitate the discussion of superconductivity, in the text we explicitly singled out the anomalous lines $S_{ij}$ and distinguished them from paramagnetic counterparts $P_{ij\sigma}$. In the present notation, the effective Hamiltonian reads

\begin{eqnarray}
  \label{eq:app:heff}
  \mathcal{H}_\mathrm{eff} = \sum \limits_l \frac{\partial \Omega_G}{\partial P_l} \hat{P}_l = \sum \limits_l \frac{N}{D} \times \left( \frac{1}{N} \frac{\partial N}{\partial P_l}  - \frac{1}{D} \frac{\partial D}{\partial P_l}  \right) \times \hat{P}_l,
\end{eqnarray}

\noindent
where $\hat{P}_l \equiv \hat{o}_1 \hat{o}_2$ and $P_l = \langle \Psi_0 | \hat{P}_l | \Psi_0\rangle$. 

\subsection{Auxiliary identity: representation of expectation values}

First we prove a useful auxiliary statement. Let us consider the expectation value

\begin{eqnarray}
  \label{eq:2}
  F \equiv \langle \Psi_0| \hat{o}_0 \cdot \hat{o}_1 \cdot \ldots \cdot \hat{o}_{2N} \cdot \hat{o}_{2N + 1} | \Psi_0 \rangle,
\end{eqnarray}

\noindent
where $N > 0$ and $\hat{o}_i$ denote either fermionic creation of annihilation operators (the factor $2N$ has been introduced in the above expression to ensure that the total number of operators is even). Additionally, we define

\begin{eqnarray}
  \label{eq:1}
  G \equiv \langle \Psi_0| \hat{o}_1 \cdot \ldots \cdot \hat{o}_{2N} | \Psi_0 \rangle.
\end{eqnarray}

\noindent
The latter is similar to $F$, but the outermost operators $\hat{o}_0$ and $\hat{o}_{2N + 1}$ have been removed.

We show next that

\begin{eqnarray}
  \label{eq:theorem}
  F =& \langle \Psi_0| \hat{o}_0 \hat{o}_{2N+1} | \Psi_0\rangle \times G + \langle \Psi_0| \hat{o}_0 \cdot \hat{G} \cdot \hat{o}_{2N + 1} |\Psi_0\rangle - \nonumber\\&- \langle \Psi_0| \hat{o}_0 \hat{o}_{2N+1} | \Psi_0\rangle \times \langle \Psi_0| \hat{G}|\Psi_0\rangle
\end{eqnarray}

with

\begin{eqnarray}
  \label{eq:def_hat_G}
  \hat{G} \equiv \sum \limits_l \frac{\partial G}{\partial P_l} \times \hat{P}_l.
\end{eqnarray}

\noindent
Once again, $l$ enumerates operator products of the form  $\hat{P}_l = \hat{o}_i \cdot \hat{o}_j$.

We start the proof by Wick decomposition of the expression for $F$:

\begin{eqnarray}
  \label{eq:contraction_F}
  F = & \langle \Psi_0| \hat{o}_0 \cdot \hat{o}_{2N + 1} | \Psi_0 \rangle \times \langle \Psi_0| \hat{o}_1 \cdot \ldots \cdot \hat{o}_{2N} | \Psi_0 \rangle + \nonumber \\ &\sum \limits_{{i, j = 1, \ldots, 2N \atop i < j}} \Big(\langle \Psi_0 | \hat{o}_0 \hat{o}_i |\Psi_0\rangle \times \langle \Psi_0 | \hat{o}_j \hat{o}_{2N + 1} |\Psi_0\rangle -\nonumber\\&- \langle \Psi_0 | \hat{o}_0 \hat{o}_j |\Psi_0\rangle \times \langle \Psi_0 | \hat{o}_i \hat{o}_{2N + 1} |\Psi_0\rangle\Big) \times \nonumber \\ & \times \langle \hat{o}_1 \cdot \ldots \cdot \hat{o}_{i - 1} \cdot \hat{o}_{i + 1} \cdot \ldots \cdot \hat{o}_{j-1} \hat{o}_{j+1} \cdot \ldots \cdot \hat{o}_{2N}\rangle \times (-1)^{j - i + 1}.
\end{eqnarray}

\noindent
The first term in equation~(\ref{eq:contraction_F}) comes from contraction of $\hat{o}_0$ and $\hat{o}_{2N + 1}$ operators, and all possible contractions of remaining terms (the latter are simply equal to the expectation value $\langle \Psi_0| \hat{o}_1 \cdot \ldots \cdot \hat{o}_{2N} | \Psi_0 \rangle$). The second term originates from contraction of $\hat{o}_0$ and $\hat{o}_{2N + 1}$ with operators from the range $\hat{o}_1, \ldots, \hat{o}_{2N}$ (and all other possible contractions of remaining terms). Note that there are two contributions in bracket of the second term due to two possible contractions of the operators $\hat{o}_0$ and $\hat{o}_{2N + 1}$ with $\hat{o}_i$ and $\hat{o}_j$. The sign $(-1)^{j - i + 1}$ arises as a consequence of moving $\hat{o}_i$ and $\hat{o}_j$ to $\hat{o}_0$ and $\hat{o}_{2N + 1}$, respectively.

Now, we point out that (for $i < j$)

\begin{eqnarray}
  \label{eq:def_G}
  G = & \langle \Psi_0| \hat{o}_i \hat{o}_j |\Psi_0\rangle \times (-1)^{j - i + 1} \times \nonumber\\ &\times \langle \hat{o}_1 \cdot \ldots \cdot \hat{o}_{i - 1} \cdot \hat{o}_{i + 1} \cdot \ldots \cdot \hat{o}_{j-1} \hat{o}_{j+1} \cdot \ldots \cdot \hat{o}_{2N}\rangle  + \nonumber \\& + \mbox{all possible contractions \textit{not} involving}\, \langle \Psi_0| \hat{o}_i \hat{o}_j |\Psi_0\rangle.
\end{eqnarray}

\noindent
There is no summation over $i$ and $j$ indices in equation~(\ref{eq:def_G}), but the latter may be chosen arbitrarily as long as $i < j$. If we treat expectation values $\langle \Psi_0| \hat{o}_i \hat{o}_j |\Psi_0\rangle$ with different sets of indices $\{i, j\}$ as independent variables, we get

\begin{eqnarray}
  \label{eq:3}
  &\frac{\partial G}{\partial \langle\Psi_0| \hat{o}_i \hat{o}_j |\Psi_0\rangle} = (-1)^{j - i + 1} \times \nonumber\\&\times \langle \hat{o}_1 \cdot \ldots \cdot \hat{o}_{i - 1} \cdot \hat{o}_{i + 1} \cdot \ldots \cdot \hat{o}_{j-1} \hat{o}_{j+1} \cdot \ldots \cdot \hat{o}_{2N}\rangle .
\end{eqnarray}

\noindent
The last equality follows from the observation that $\langle\Psi_0| \hat{o}_i \hat{o}_j |\Psi_0\rangle$ appears \emph{exactly} once in equation~(\ref{eq:def_G}) as a coefficient of the first term. We thus arrive at a more compact formula

\begin{eqnarray}
  \label{eq:contraction_F_2}
  F = & \langle \Psi_0| \hat{o}_0 \cdot \hat{o}_{2N + 1} | \Psi_0 \rangle \times \langle \Psi_0| \hat{o_1} \cdot \ldots \cdot \hat{o}_{2N} | \Psi_0 \rangle + \nonumber \\ &\sum \limits_{{i, j = 1, \ldots, 2N \atop i < j}} \Big(\langle \Psi_0 | \hat{o}_0 \hat{o}_i |\Psi_0\rangle \times \langle \Psi_0 | \hat{o}_j \hat{o}_{2N + 1} |\Psi_0\rangle - \nonumber\\&- \langle \Psi_0 | \hat{o}_0 \hat{o}_j |\Psi_0\rangle \times \langle \Psi_0 | \hat{o}_i \hat{o}_{2N + 1} |\Psi_0\rangle\Big) \times \frac{\partial G}{\partial \langle \Psi_0| \hat{o}_i \hat{o}_j |\Psi_0\rangle}.
\end{eqnarray}

\noindent
that can be further simplified by application of the Wick's theorem to the expressions in bracket of the second term:

\begin{eqnarray}
  \label{eq:contraction_F_3}
  F = & \langle \Psi_0| \hat{o}_0 \cdot \hat{o}_{2N + 1} | \Psi_0 \rangle \times \langle \Psi_0| \hat{o}_1 \cdot \ldots \cdot \hat{o}_{2N} | \Psi_0 \rangle + \nonumber \\ &\sum \limits_{{i, j = 1, \ldots, 2N \atop i < j}} \Big(\langle \Psi_0| \hat{o}_0  \hat{o}_i \hat{o}_j \hat{o}_{2N + 1}|\Psi_0\rangle -\nonumber\\&- \langle \Psi_0| \hat{o}_0 \hat{o}_{2N + 1}|\Psi_0\rangle \times \langle \Psi_0| \hat{o}_i \hat{o}_{j} |\Psi_0\rangle\Big) \times  \frac{\partial G}{\partial \langle \Psi_0| \hat{o}_i \hat{o}_j |\Psi_0\rangle}.
\end{eqnarray}

Note that different lines $\langle \Psi_0| \hat{o}_i \hat{o}_j |\Psi_0\rangle$ and $\langle\Psi_0| \hat{o}_{i'} \hat{o}_{j'} |\Psi_0\rangle$ are certainly equal if $\hat{o}_i \hat{o}_j = \hat{o}_{i'} \hat{o}_{j'}$. It is thus useful to collect the lines that are expectation values of the same operator products so that $\langle \Psi_0| \hat{o}_i \hat{o}_j |\Psi_0\rangle$ and $\langle \Psi_0| \hat{o}_{i'} \hat{o}_{j'} |\Psi_0\rangle$ are now treated as the same variable if $\hat{o}_i \hat{o}_j = \hat{o}_{i'} \hat{o}_{j'}$. The derivative over lines in equation~(\ref{eq:contraction_F_3}) should be then transformed according to the relation

\begin{eqnarray}
  \label{eq:5}
  \frac{\partial G}{\partial P_l} = \sum \limits_{{i, j = 1, \ldots, 2N;  i < j \atop \hat{o}_i \hat{o}_j = \hat{P}_l}}   \frac{\partial G}{\partial \langle \Psi_0| \hat{o}_i \hat{o}_j |\Psi_0\rangle}.
\end{eqnarray}

\noindent
The expression for $F$ can be then written in its final form

\begin{eqnarray}
  \label{eq:contraction_F_4}
  F = & \langle \Psi_0| \hat{o}_0 \cdot \hat{o}_{2N + 1} | \Psi_0 \rangle \times \langle \Psi_0| \hat{o}_1 \cdot \ldots \cdot \hat{o}_{2N} | \Psi_0 \rangle + \nonumber \\ &\sum \limits_{l} (\langle \Psi_0| \hat{o}_0  \hat{P}_l \hat{o}_{2N + 1}|\Psi_0\rangle - \langle \Psi_0| \hat{o}_0 \hat{o}_{2N + 1}|\Psi_0\rangle \times \langle \Psi_0| \hat{P}_l |\Psi_0\rangle ) \times \frac{\partial G}{\partial P_l} =  \nonumber \\ & \langle \Psi_0| \hat{o}_0 \hat{o}_{2N+1} | \Psi_0\rangle \times G + \langle \Psi_0| \hat{o}_0 \cdot \hat{G} \cdot \hat{o}_{2N + 1} |\Psi_0\rangle \nonumber - \\ & - \langle \Psi_0| \hat{o}_0 \hat{o}_{2N+1} | \Psi_0\rangle \times \langle \Psi_0| \hat{G} |\Psi_0\rangle,
\end{eqnarray}

\noindent
which completes the reasoning.

\subsection{Energy spectrum of the effective Hamiltonian}

We are now in position to relate the expression for the spectrum of the effective Hamiltonian $\mathcal{H}_\mathrm{eff}$ to the projected wave functions. We first calculate the value of the grand potential in the state with added projected quasiparticle excitation

\begin{eqnarray}
  \label{eq:projected_qp_energy}
  \Omega_{G\mathbf{k}\sigma} \equiv \frac{\langle\Psi_0| \gamma_{\mathbf{k}\sigma}P_G (\mathcal{H} - \mu \hat{N})P_G \gamma_{\mathbf{k}\sigma}^\dagger|\Psi_0\rangle}{\langle\Psi_0| \gamma_{\mathbf{k}\sigma}P_G^2 \gamma_{\mathbf{k}\sigma}^\dagger|\Psi_0\rangle}.
\end{eqnarray}

\noindent
We  can now apply the decomposition of equation~(\ref{eq:theorem}) both to the nominator and denominator of equation~(\ref{eq:projected_qp_energy}). The use of equation~(\ref{eq:theorem}) is admissible as Bogoliubov quasiparticles are linear combinations of the original creation and annihilation operators. We get

\begin{eqnarray}
  \label{eq:projected_qp_energy_1}
  \Omega_{G\mathbf{k}\sigma} =& \Big( \langle\Psi_0|\gamma_{\mathbf{k}\sigma} \gamma_{\mathbf{k}\sigma}^\dagger|\Psi_0\rangle  N + \sum \limits_l \frac{\partial N}{\partial P_l} \langle\Psi_0|\gamma_{\mathbf{k}\sigma}  \hat{P}_l  \gamma_{\mathbf{k}\sigma}^\dagger|\Psi_0\rangle - \nonumber \\ & -  \sum \limits_l \frac{\partial N}{\partial P_l} \langle\Psi_0|\gamma_{\mathbf{k}\sigma}  \gamma_{\mathbf{k}\sigma}^\dagger|\Psi_0\rangle  \langle\Psi_0|\hat{P}_l|\Psi_0\rangle}   \Big)   \times   \Big(      {\langle\Psi_0|\gamma_{\mathbf{k}\sigma} \gamma_{\mathbf{k}\sigma}^\dagger|\Psi_0\rangle  D + \nonumber \\ &+ \sum \limits_l \frac{\partial D}{\partial P_l} \langle\Psi_0|\gamma_{\mathbf{k}\sigma}  \hat{P}_l  \gamma_{\mathbf{k}\sigma}^\dagger|\Psi_0\rangle - \nonumber \\ &-  \sum \limits_l \frac{\partial D}{\partial P_l} \langle\Psi_0|\gamma_{\mathbf{k}\sigma}  \gamma_{\mathbf{k}\sigma}^\dagger|\Psi_0\rangle  \langle\Psi_0|\hat{P}_l|\Psi_0\rangle \Big)^{-1}.
\end{eqnarray}

\noindent
Now note that, if $\gamma_{\mathbf{k}\sigma}$ creates a quasi-particle excitation in the state $\mathbf{k}\sigma$, this state must be unoccupied in the first place, thus $\langle\Psi_0|\gamma_{\mathbf{k}\sigma} \gamma_{\mathbf{k}\sigma}^\dagger|\Psi_0\rangle = \langle\Psi_0|1 - \gamma^\dagger_{\mathbf{k}\sigma} \gamma_{\mathbf{k}\sigma}|\Psi_0\rangle =  1$. Taking this into account and reorganizing equation~(\ref{eq:projected_qp_energy_1}), we arrive at

\begin{eqnarray}
  \label{eq:projected_qp_energy_2}
  \Omega_{G\mathbf{k}\sigma} = \frac{N}{D} \times \frac{1 + \sum \limits_l \frac{1}{N}\frac{\partial N}{\partial P_l} \langle\Psi_0|\gamma_{\mathbf{k}\sigma}  \hat{P}_l \gamma_{\mathbf{k}\sigma}^\dagger|\Psi_0\rangle - \sum \limits_l \frac{1}{N}\frac{\partial N}{\partial P_l} \langle\Psi_0|\hat{P}_l|\Psi_0\rangle}{1 + \sum \limits_l \frac{1}{D}\frac{\partial D}{\partial P_l} \langle\Psi_0|\gamma_{\mathbf{k}\sigma}  \hat{P}_l \gamma_{\mathbf{k}\sigma}^\dagger|\Psi_0\rangle - \sum \limits_l \frac{1}{D}\frac{\partial D}{\partial P_l} \langle\Psi_0|\hat{P}_l|\Psi_0\rangle}.
\end{eqnarray}

\noindent
The denominator can be expanded in the Taylor series with respect to sums over density matrix elements. As is detailed in the following subsection, the series should converge rapidly for the case of Gutzwiller wave function. By keeping the terms up to the first expansion order, we get

\begin{eqnarray}
  \label{eq:projected_qp_energy_3}
    \Omega_{G\mathbf{k}\sigma} \approx &\frac{N}{D} + \frac{N}{D} \sum \limits_l \left( \frac{1}{N}\frac{\partial N}{\partial P_l} - \frac{1}{D}\frac{\partial D}{\partial P_l}\right)  \times \langle\Psi_0|\gamma_{\mathbf{k}\sigma} \cdot \hat{P}_l \cdot \gamma^\dagger_{\mathbf{k}\sigma}|\Psi_0\rangle - \nonumber \\ & - \frac{N}{D} \sum \limits_l \left( \frac{1}{N}\frac{\partial N}{\partial P_l} - \frac{1}{D}\frac{\partial D}{\partial P_l}\right)  \times \langle\Psi_0|\hat{P}_l|\Psi_0\rangle.
\end{eqnarray}

\noindent

The latter expression can be reorganized with the use of equation~(\ref{eq:app:heff})

\begin{eqnarray}
  \label{eq:projected_qp_energy_final}
  \Omega_{G\mathbf{k}\sigma} \approx&  \Omega_G + \langle\Psi_0|\gamma_{\mathbf{k}\sigma}\mathcal{H}_\mathrm{eff} \gamma_{\mathbf{k}\sigma}^\dagger|\Psi_0\rangle - \langle\Psi_0|\mathcal{H}_\mathrm{eff}|\Psi_0\rangle = \nonumber \\ &\Omega_G + \sum_{\mathbf{k}'\sigma'} \epsilon^\mathrm{eff}_{\mathbf{k}'\sigma'}\left(\langle\Psi_0| \gamma_{\mathbf{k}\sigma} \gamma_{\mathbf{k}'\sigma'}^\dagger \gamma_{\mathbf{k}'\sigma'} \gamma_{\mathbf{k}\sigma}^\dagger|\Psi_0\rangle - \langle\Psi_0| \gamma_{\mathbf{k}'\sigma'}^\dagger \gamma_{\mathbf{k}'\sigma'}|\Psi_0\rangle\right) = \nonumber \\ & = \Omega_G + \epsilon_{\mathbf{k}\sigma}^\mathrm{eff},
\end{eqnarray}

\noindent
where $\epsilon_{\mathbf{k}\sigma}^\mathrm{eff}$ is the quasi-particle dispersion obtained by diagonalization of the
effective Hamiltonian $\mathcal{H}_\mathrm{eff}$. This completes the reasoning.

\subsection{Justification of the denominator expansion for the case of Gutzwiller wave function}

Here we justify the transition form equation~(\ref{eq:projected_qp_energy_2}) to equation~(\ref{eq:projected_qp_energy_3}). The argument is not general and bases on the specific form of Gutzwiller wave function.

First, we develop a formal  expansion for the denominator of the right-hand-side of equation~(\ref{eq:projected_qp_energy_2}) in the powers of the parameter $x$. To do that, we return to its original form $\langle \Psi_0| \gamma_{\mathbf{k}\sigma} P_G^2 \gamma_{\mathbf{k}\sigma}^\dagger|\Psi_0\rangle/D = \langle \Psi_0| \gamma_{\mathbf{k}\sigma} P_G^2 \gamma_{\mathbf{k}\sigma}^\dagger|\Psi_0\rangle/\langle \Psi_0| P_G^2 |\Psi_0\rangle$. By making use of the identity  $P_G^2 = \prod_i (1 + x \cdot \hat{d}_i^\mathrm{HF})$ with $\hat{d}_i^\mathrm{HF} = \hat{n}^\mathrm{HF}_{i_\uparrow} \hat{n}^\mathrm{HF}_{i\downarrow}$, and collecting the terms proportional to the same powers of $x$, we arrive at

\begin{eqnarray}
  \label{eq:denominator_expanded}
  &\frac{\langle \Psi_0| \gamma_{\mathbf{k}\sigma} P_G^2 \gamma_{\mathbf{k}\sigma}^\dagger|\Psi_0\rangle}{\langle \Psi_0| P_G^2 |\Psi_0\rangle}  = \sum_{\boldsymbol{\delta}} \exp(i \mathbf{k} \boldsymbol{\delta}) \times \nonumber\\&\times \frac{\sum \limits_{k = 0}^\infty \frac{x^k}{k!} \sideset{}{_{}^{'}} \sum \limits_{l_1, \ldots, l_k} \langle\Psi_0| \gamma_{i\sigma} \cdot   \hat{d}^\mathrm{HF}_{l_1}\cdot\ldots\cdot\hat{d}^\mathrm{HF}_{l_k} \cdot  \gamma^\dagger_{i + \boldsymbol{\delta},\sigma}|\Psi_0\rangle}{\sum \limits_{k = 0}^\infty \frac{x^k}{k!} \sideset{}{_{}^{'}} \sum \limits_{l_1, \ldots, l_k} \langle\Psi_0| \hat{d}^\mathrm{HF}_{l_1}\cdot\ldots\cdot\hat{d}^\mathrm{HF}_{l_k}|\Psi_0\rangle},
\end{eqnarray}

\noindent
where primes mean that the summation is performed over sets of unique lattice indices (i.e. $l_\alpha \neq l_\beta$ for $\alpha \neq \beta$). The right-hand side of equation~(\ref{eq:denominator_expanded}) can be now expanded with the use of Wick's theorem. We get

\begin{eqnarray}
  \label{eq:denominator_expanded_final}
  &\frac{\langle \Psi_0| \gamma_{\mathbf{k}\sigma} P_G^2 \gamma_{\mathbf{k}\sigma}^\dagger|\Psi_0\rangle}{\langle \Psi_0| P_G^2 |\Psi_0\rangle}  = \sum_{\boldsymbol{\delta}} \exp(i \mathbf{k} \boldsymbol{\delta})  \langle\Psi_0| \gamma_{i\sigma} \gamma^\dagger_{i + \boldsymbol{\delta}, \sigma} |\Psi_0\rangle + \nonumber \\ &+ \frac{x^2}{2!}\sum_{\boldsymbol{\delta}} \exp(i \mathbf{k} \boldsymbol{\delta})  \sum\limits_{l_1 \neq l_2} \langle \Psi_0| \gamma_{i\sigma} \hat{d}_{l_1}^\mathrm{HF} \hat{d}_{l_2}^\mathrm{HF } \gamma^\dagger_{i + \boldsymbol{\delta}, \sigma}|\Psi_0\rangle_c  + o(x^2),
\end{eqnarray}

\noindent
where the subscript ``$c$'' means that only the connected diagrams should be included. The disconnected part has been canceled by denominator contribution. Note that all the terms proportional to $x$ vanish, i.e. $\langle\Psi_0| \hat{d}_{l_1}^\mathrm{HF} |\Psi_0\rangle = 0$ and $\langle\Psi_0| \gamma_{i\sigma} \hat{d}_{l_1}^\mathrm{HF} \gamma_{i + \boldsymbol{\delta}, \sigma}|\Psi_0\rangle = 0$.  As we argued above, the state $\mathbf{k}\sigma$ must be unoccupied, hence $\sum_{\boldsymbol{\delta}} \exp(i\mathbf{k}\boldsymbol{\delta}) \langle\Psi_0| \gamma_{i\sigma}\gamma^\dagger_{i + \boldsymbol{\delta},\sigma}|\Psi_0\rangle = \langle\Psi_0|\gamma_{\mathbf{k}\sigma} \gamma^\dagger_{\mathbf{k}\sigma}|\Psi_0\rangle = 1$. By comparing equation~(\ref{eq:denominator_expanded_final}) with the denominator of equation~(\ref{eq:projected_qp_energy_2}), we observe that

\begin{eqnarray}
  \label{eq:denominator_expanded_comparison}
  &\sum \limits_l \frac{1}{D}\frac{\partial D}{\partial P_l} \langle\Psi_0|\gamma_{\mathbf{k}\sigma} \cdot \hat{P}_l \cdot \gamma_{\mathbf{k}\sigma}^\dagger|\Psi_0\rangle - \sum \limits_l \frac{1}{D}\frac{\partial D}{\partial P_l} \langle\Psi_0|\hat{P}_l|\Psi_0\rangle = \nonumber \\  &=
  \frac{x^2}{2!} \sum_{\boldsymbol{\delta}} \exp(i \mathbf{k} \boldsymbol{\delta})  \sum \limits_{l_1, l_2} \langle\Psi_0| \gamma_{i\sigma} \cdot   \hat{d}^\mathrm{HF}_{l_1}\cdot\hat{d}^\mathrm{HF}_{l_2} \cdot  \gamma^\dagger_{i + \boldsymbol{\delta},\sigma}|\Psi_0\rangle_c + \nonumber\\ &+ o(x^2).
\end{eqnarray}

\noindent
The expansion performed in the process of transforming  equation~(\ref{eq:projected_qp_energy_2}) into equation~(\ref{eq:projected_qp_energy_3}) is thus valid to the order $o(x^2)$.

\section{Evaluation of the $\mathbf{k}$-space diagrammatic sums}
\label{appendix:graphs}

Here we present the details of $\mathbf{k}$-space diagrammatic expansion, constituting the basis for the $\mathbf{k}$-DE-GWF approach.

First, a methodological remark is in order. The naive formulation of the diagrammatic expansion, based on Fourier transformation of real-space Wick-factorized sums (\ref{eq:t11})-(\ref{eq:i4}) would lead to complicated $\mathbf{k}$-space expressions due to summation restrictions over internal vertex positions $\{l_\alpha\}$ (indicated by primes over summation symbols). By using $T^{11}_{ij}$ as an example, we now show how this problem is eliminated by proper redefinition of lines appearing in the Wick's decomposition; all other sums can be transformed in an analogous way. Namely, for $i \neq j$, we have

\begin{eqnarray}
  T^{11}_{ij} &= \sum \limits_{k=0}^{\infty} \frac{x^k}{k!} \sideset{}{_{}^{'}} \sum_{l_1, \ldots, l_k} \langle \Psi_0| c^\dagger_{i\uparrow} c_{j\uparrow} \hat{d}^\mathrm{HF}_{l_1} \ldots \hat{d}^\mathrm{HF}_{l_k}|\Psi_0\rangle_{c} = \nonumber \\ &= \sum \limits_{k=0}^{\infty} \frac{x^k}{k!}  \sum_{l_1, \ldots, l_k} \langle \Psi_0| c^\dagger_{i\uparrow} c_{j\uparrow} \hat{d}_{l_1} \ldots \hat{d}_{l_k}|\Psi_0\rangle_{c}^{'},
\label{app:eq:representation_of_t11}
\end{eqnarray}


\noindent
where now summation restrictions over $\{l_\alpha\}$ are lifted and the full operators $\hat{d}_i = \hat{n}_{i\uparrow}\hat{n}_{i\downarrow}$ are used instead of their shifted versions $\hat{d}_i^\mathrm{HF} = (\hat{n}_{i\uparrow} - n_\uparrow^0) (\hat{n}_{i\downarrow} - n^0_\downarrow)$. The prime over the expectation value on the right-hand-side of equation~(\ref{app:eq:representation_of_t11}) means that modified paramagnetic lines $\langle\Psi_0| c^\dagger_{i\sigma} c_{j\sigma} |\Psi_0\rangle \rightarrow \langle\Psi_0| c^\dagger_{i\sigma} c_{j\sigma} |\Psi_0\rangle - \delta_{ij} \langle\Psi_0| c^\dagger_{i\sigma} c_{j\sigma} |\Psi_0\rangle$ and $\langle\Psi_0| c_{i\sigma} c^\dagger_{j\sigma} |\Psi_0\rangle \rightarrow \langle\Psi_0| c_{i\sigma} c^\dagger_{j\sigma} |\Psi_0\rangle - \delta_{ij} \langle\Psi_0| c_{i\sigma} c^\dagger_{j\sigma} |\Psi_0\rangle$  should be used in the Wick's expansion \cite{BunemannEPL2012}. To show that these two representations are equivalent, we point out the following. The effect of using the shifted version of the double occupancy operators $\hat{d}_i^\mathrm{HF}$ in the original representation of $T^{11}_{ij}$ is to eliminate Hartree-bubbles attached to internal vertices. Due to redefinition of the paramagnetic lines, however, all Hartree-bubbles vanish by construction if $\hat{d}_i^\mathrm{HF}$ are substituted with $\hat{d}_i$. Note that all diagrams without Hartree-bubbles remain the same as would come out from the original expression since the modification is limited to local correlation functions. Moreover, the summation restrictions can be lifted because diagrammatic contributions involving two terms located on the same lattice site add up to zero. To illustrate the last point, we consider an exemplary term contributing to $T^{11}_{ij}$ with an equal internal and external lattice index, written schematically as $\langle \Psi_0| \ldots \cdot \hat{c}^\dagger_{i\uparrow}\cdot  \hat{d}_i \cdot \ldots |\Psi_0\rangle_c^{'}$, where dots represent products of other operators. In the process of Wick-decomposition of this sum we encounter (i) terms containing the local contraction between $\hat{d}_i$ and $c_{i\uparrow}^\dagger$ operator $\langle \Psi_0| \wick{ \ldots \cdot \c1 {c_{i\uparrow}^\dagger} \cdot c^\dagger_{i\uparrow} \c1 {c_{i\uparrow}} c^\dagger_{i\downarrow} c_{i\downarrow}  \cdot \ldots }|\Psi_0\rangle_c^{'}$, and (ii) the terms, where $\hat{d}_i$ and $c_{i\uparrow}$ have been contracted with some other operators. The diagrams of type (i) vanish as a consequence of using shifted lines. The diagrams from the class (ii), on the other hand, always contain pairs of contractions that can be represented as

\begin{eqnarray}
\langle\Psi_0| \wick{ \ldots \cdot \c1{c_{m\uparrow}} \ldots \cdot  \c1 {c_{i\uparrow}^\dagger} \cdot \c1{c^\dagger_{i\uparrow}}  c_{i\uparrow} c^\dagger_{i\downarrow} c_{i\downarrow} \cdot \ldots \cdot \c1{c_{n\uparrow}} \cdot \ldots }|\Psi_0\rangle_c^{'}
\end{eqnarray}

\noindent
and

\begin{eqnarray}
\langle\Psi_0| \wick{ \ldots \cdot \c1{c_{m\uparrow}} \ldots \cdot  \c2 {c_{i\uparrow}^\dagger} \cdot \c1{c^\dagger_{i\uparrow}}  c_{i\uparrow} c^\dagger_{i\downarrow} c_{i\downarrow} \cdot \ldots \cdot \c2{c_{n\uparrow}} \cdot \ldots }|\Psi_0\rangle_c^{'}.
\end{eqnarray}

\noindent
These two terms are equal in regard of absolute values, but differ in sign and thus add up to zero, which concludes the reasoning.

Whereas the expressions for the two-site sums are, strictly speaking, defined only for $i \neq j$, it is convenient do set $T^{11}_{ii} \equiv \sum_{k=0}^{\infty} \frac{x^k}{k!}  \sum_{l_1, \ldots, l_k} \langle\Psi_0| c^\dagger_{i\uparrow} c_{i\uparrow} \hat{d}_{l_1} \ldots \hat{d}_{l_k}|\Psi_0\rangle_{c}^{'}$. The value of $T^{11}_{ii}$ is physically irrelevant as it does not enter any of the functionals used in our scheme, but adopting this particular form leads to simplification of the $\mathbf{k}$-space expressions. 

Now, the real-space Wick's decomposition of the expectation values (\ref{eq:t11})-(\ref{eq:i4}) in the modified form, explained above, has been performed for the case of (i) the paramagnetic phase with only normal lines $\tilde{P}_{ij\sigma} \equiv P_{ij\sigma} - \delta_{ij} n^0_\sigma$ included (note the shift due to line redefinition), and (ii) $d$-wave SC state allowing for both normal $\tilde{P}_{ij\sigma}$ and anomalous terms $S_{ij}$. Inversion symmetry was assumed in the process, which resulted in technical simplifications. Subsequently, the isomorphic graphs were collected and combinatorical factors were calculated. The Nauty library \cite{McKayJSymbComput2014} was utilized for low-level graph-theory operations. Finally, the resultant diagrammatic sums have been Fourier transformed to the $\mathbf{k}$-space and the local wave vector conservation (modulo reciprocal lattice vector) has been exploited to eliminate excess $\mathbf{k}$-space integrals. At this stage, the $\mathbf{k}$-space sums of the form $T^{\alpha\beta}(\mathbf{p}) \equiv \sum_{\boldsymbol{\delta}} T^{\alpha\beta}_{i+\boldsymbol{\delta}, i} \exp(i \mathbf{p} \boldsymbol{\delta})$ and $S^{22}(\mathbf{k}) \equiv \sum_{\boldsymbol{\delta}} S^{22}_{i+\boldsymbol{\delta}, i} \exp(i \mathbf{p} \boldsymbol{\delta})$ can be readily expressed in terms of integrals involving the $\mathbf{k}$-space lines $\tilde{P}_{\mathbf{k}} \equiv \sum_{\boldsymbol{\delta}} P_{i+\boldsymbol{\delta}, i \sigma} \exp(i \mathbf{k}\boldsymbol{\delta}) - n^0_\sigma$ and $S_{\mathbf{k}} \equiv \sum_{\boldsymbol{\delta}} S_{i+\boldsymbol{\delta}, i} \exp(i \mathbf{k}\boldsymbol{\delta})$ (since we do not consider magnetism, the spin symbol $\sigma$ has been omitted in the definition of $\tilde{P}_\mathbf{k}$). Note that we have not imposed the condition $\boldsymbol{\delta} \neq \mathbf{0}$ in the definition of the $\mathbf{k}$-space expressions. Additionally, we have reserved the symbol $\mathbf{p}$ to denote the external wave vectors (flowing through entire diagrammatic sum) that will be contrasted with dummy internal ones $\mathbf{k}_\alpha$ over which integration is performed.

\begin{figure*}[h]
\includegraphics[width=0.95\textwidth]{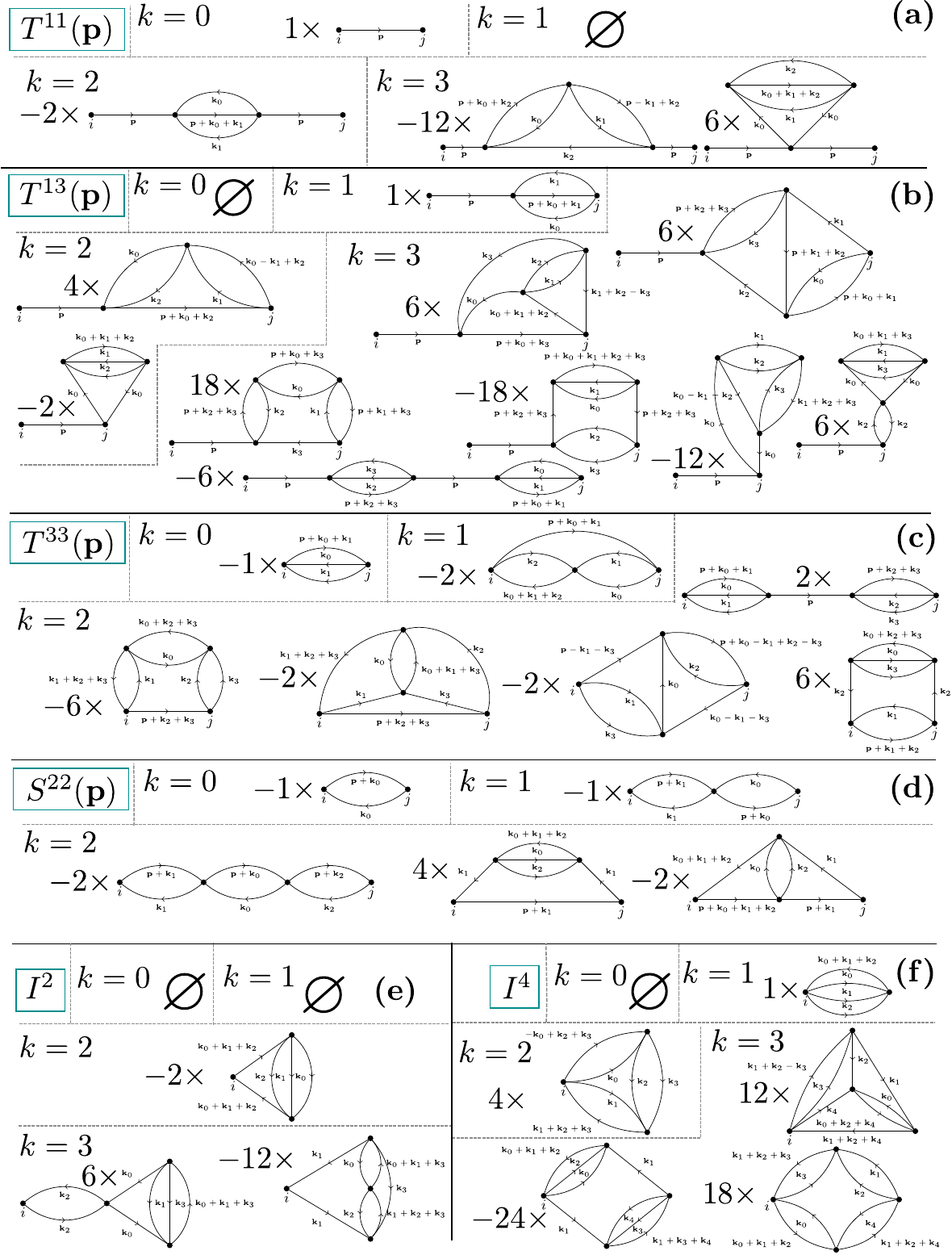}
    \caption{The paramagnetic $\mathbf{k}$-space diagrams contributing to the variational energy up to the order ($k \leq 3$). Some $k = 3$ diagram classes are not displayed. The lines have been assigned the combinations of external and internal wave vectors ($\mathbf{p}$ and $\mathbf{k}_\alpha$, respectively). The external vertices are labeled by indices $i$ and $j$. The numbers next to the diagrams are combinatorical factors.}
    \label{fig:diagrams}
\end{figure*}

In figure~\ref{fig:diagrams} we display the paramagnetic diagrams contributing to the variational energy up to the third expansion order $k$. In that case, each line of the diagram corresponds to the $\mathbf{k}$-space expression $\tilde{P}_{\mathbf{k}}$ with the appropriate combination of wave vectors written near the line. The numbers next to the graphs are combinatorical coefficients. The symbols $i$ and $j$ near the vertices mark the external vertices. For illustration, we write down the explicit expression for $T^{11}(\mathbf{p})$ in the paramagnetic phase up to the second expansion order

\begin{eqnarray}
  T^{11}(\mathbf{p}) &= \tilde{P}_\mathbf{p} + \frac{x^2}{2!} \times (-2) \times \tilde{P}_\mathbf{p}^2 \times \nonumber\\& \times\int \frac{d^2\mathbf{k}_0}{(2\pi)^2} \frac{d^2\mathbf{k}_1}{(2\pi)^2} \tilde{P}_{\mathbf{p} + \mathbf{k}_0 + \mathbf{k}_1} \tilde{P}_{-\mathbf{k}_0} \tilde{P}_{-\mathbf{k}_1}  + o(x^2),
\label{app:eq:explicit_expression_t11}
\end{eqnarray}

\noindent
where all integrals are performed over the first Brillouin zone (cf. figure~\ref{fig:diagrams}(a)).  The one-site sums do not carry external wave vector $\mathbf{p}$ and have simpler structure, e.g.,

\begin{eqnarray}
  &I^2 = \frac{x^2}{2!} \times (-2) \times \int \frac{d^2\mathbf{k}_0}{(2\pi)^2} \times \nonumber\\&\times \int \frac{d^2\mathbf{k}_1}{(2\pi)^2} \int \frac{d^2\mathbf{k}_2}{(2\pi)^2} \tilde{P}_{-\mathbf{k}_0 - \mathbf{k}_1 - \mathbf{k}_2}^2 \tilde{P}_{\mathbf{k}_0} \tilde{P}_{\mathbf{k}_1} \tilde{P}_{\mathbf{k}_2} + o(x^2).
\end{eqnarray}

\noindent
Having computed all diagrammatic contributions to desired order in this way, one can retrieve the variational energy components (\ref{eq:kinetic_energy_contributio_diag})-(\ref{eq:exchange_energy_contributio_diag}) by Fourier transforming to real-space again. For $\boldsymbol{\delta}\neq \mathbf{0}$ we get explicitly $T^{\alpha\beta}_{i+\boldsymbol{\delta}, i} = \int \frac{d^2\mathbf{p}}{(2\pi)^2} \exp(-i \mathbf{p}\boldsymbol{\delta}) T^{\alpha\beta}(\mathbf{p}$) and $S^{22}_{i+\boldsymbol{\delta}, i} = \int \frac{d^2\mathbf{p}}{(2\pi)^2} \exp(-i \mathbf{p}\boldsymbol{\delta}) S^{22}(\mathbf{p})$.

{
  \begin{table}[h!]
    \caption{Number of diagrams contributing to various expectation values ($N_\mathrm{diag}$) and the corresponding $\mathbf{k}$-space  integral dimensions ($D$) in the paramagnetic (PM) and superconducting ($d$-SC) phases as a function of expansion order $k$. Here $N_\mathrm{total}$ denotes the total number diagrams of given order.}
    \label{table:diagram_numbers}
    \centering
  \begin{tabular}{ c  c  c c c c c  c c c c c}
    \hline\hline
   \multicolumn{2}{c}{} & \multicolumn{5}{c  }{PM phase} & \multicolumn{5}{c}{$d$-SC phase} \\
   \multicolumn{2}{c}{} & \multicolumn{5}{c  }{Expansion order, $k$} & \multicolumn{5}{c}{Expansion order, $k$} \\
   \multicolumn{2}{c}{} & $0$ & $1$ & $2$ & $3$ & $4$ & $0$ & $1$ & $2$ & $3$ & $4$ \\
    \hline \hline 
    $T^{11}$ &  $N_\mathrm{diag}$ & 1 & 0 & 1 & 2 & 7 & 1 & 0 & 8 & 40 & 644 \\
                        & $D$ & 2 & N/A  & 6  & 8  & 10  & 2  & N/A  & 6  & 8  & 10  \\
    $T^{13}$ & $N_\mathrm{diag}$ & 0 & 1 & 2 & 7 & 27 & 0 & 4 & 20 & 322 & 3262 \\
                            & $D$ & N/A & 6 & 8 & 10 & 12 & N/A & 6 & 8 & 10 & 12 \\
    $T^{33}$ & $N_\mathrm{diag}$ & 1 & 1 & 5 & 20 & 91 & 2 & 4 & 133 & 1238 & 27187 \\
                            & $D$ & 6 & 8 & 10 & 12 & 14 & 6 & 8 & 10 & 12 & 14 \\
    $S^{22}$ & $N_\mathrm{diag}$ & 1 & 1 & 3 & 8 & 34 & 2 & 4 & 36 & 306 & 4388 \\
                            & $D$ & 2 & 4 & 6 & 8 & 12 & 2 & 4 & 6 & 8 & 12 \\
    $I^2$ &  $N_\mathrm{diag}$ & 0 & 0 & 1 & 2 & 7 & 0 & 0 & 6 & 28 & 393 \\
                            & $D$ & N/A & N/A & 6 & 8 & 10 & N/A & N/A & 6 & 8 & 10 \\
    $I^4$ &  $N_\mathrm{diag}$ & 0 & 1 & 1 & 3 & 12 & 0 & 3 & 6 & 103 & 888 \\
                            & $D$ & N/A & 6 & 8 & 10 & 12 & N/A & 6 & 8 & 10 & 12 \\
                        & $N_\mathrm{total}$ & 3 & 4 & 13 & 42  & 178  & 5  & 15  & 209  & 2037 & 36762  \\
    \hline\hline
  \end{tabular}
\end{table}
}

In table \ref{table:diagram_numbers} we list the number ($N_\mathrm{diag}$) of generated diagrams, contributing to different expectation values, in the paramagnetic and $d$-SC phases up to the fourth expansion order $k$. Additionally, we display the dimensions $D$ of $\mathbf{k}$-space integrals that arise from diagrammatic contributions to the variational energy functional $E_G$ and its derivatives over lines. The total graph count $N_\mathrm{total}$ is given in the last row. Note that $N_\mathrm{total}$ rapidly increases in the SC state which provides a practical limitation for the $\mathbf{k}$-DE-GWF method. Here we have been able to include the diagrams up to the third expansion order ($k \leq 3$) in the SC state ($\sim 2000$ graphs resulting in at most $12$-dimensional integrals). The $\mathbf{k}$-space Monte-Carlo integration has been performed using Suave algorithm as implemented in the Cuba library.

\section{Diagrammatic representation of the spectral function moments}
\label{appendix:representation_of_m1}

Here we develop diagrammatic representation of the two leading moments of the spectral function $\mathcal{M}_0(\mathbf{k})$ and $\mathcal{M}_1(\mathbf{k})$, suitable for evaluation within the $\mathbf{k}$-DE-GWF method. To relate the latter to the ground-state expectation values that can be calculated diagramatically, we make use of the Lehmann representation of zero-temperature time-ordered Green's functions

\begin{eqnarray}
  \label{eq:lehmann_representation}
  G(\mathbf{k}, \omega) &\equiv -i \int \limits_{-\infty}^{\infty} dt \mathrm{e}^{i \omega t} \langle \mathcal{T} c_{\mathbf{k}\sigma}(t) c^\dagger_{\mathbf{k}\sigma} \rangle = \int \limits_{-\infty}^{\infty} dE \frac{\mathcal{A}(\mathbf{k}, E)}{\omega - E + i\epsilon\cdot\mathrm{sgn}(w)},
\end{eqnarray}

\noindent
where $\mathcal{T}$ is time-ordering operator and $\epsilon$ denotes an infinitesimal positive term. By Fourier-transforming over $\omega$ to get $G(\mathbf{k}, t) = \int \frac{d\omega}{2\pi} G(\mathbf{k}, \omega) \exp(- i \omega t)$, and taking the time derivatives of both the Green's function definition and its Lehmann representation, we arrive at

\begin{eqnarray}
  \label{eq:app:M0}
  \lim \limits_{t \rightarrow 0^{-}} G(\mathbf{k}, t) = i \langle c^\dagger_{\mathbf{k}\sigma}c_{\mathbf{k}\sigma} \rangle = i \int \limits_{-\infty}^0 d\omega \mathcal{A}(\mathbf{k}, \omega),
\end{eqnarray}

\noindent
and

\begin{eqnarray}
  \label{eq:app:M1}
  \lim \limits_{t \rightarrow 0^{-}} \partial_t G(\mathbf{k}, t) &= - \langle c^\dagger_{\mathbf{k}\sigma} \cdot [\mathcal{H} - \mu \hat{N}, c_{\mathbf{k}\sigma}] \rangle = \int \limits_{-\infty}^0 d\omega \omega \mathcal{A}(\mathbf{k}, \omega).
\end{eqnarray}

\noindent
The zero as the upper limit in above expression follows from closing the integration contour over $E$ variable in the upper complex half-plane. Another way to derive equations~(\ref{eq:app:M0})-(\ref{eq:app:M1}) is to make use of the explicit form of the zero-temperature spectral function

\begin{eqnarray}
  \mathcal{A}(\mathbf{k}, \omega) &= \sum \limits_n \langle 0 | c^\dagger_{\mathbf{k}\sigma}|n\rangle\langle n| c_{\mathbf{k}\sigma}|0\rangle \delta(\omega + E_n - E_0) + \nonumber\\& + \sum \limits_n \langle 0 | c_{\mathbf{k}\sigma}|n\rangle\langle n| c^\dagger_{\mathbf{k}\sigma}|0\rangle \delta(\omega - E_n + E_0)
\end{eqnarray}

\noindent
and evaluate the moments directly. Here $|n\rangle$ and $E_n$ are exact eigenstates of $\mathcal{H} - \mu\hat{N}$ and corresponding energies, respectively.

The relations (\ref{eq:app:M0}) and (\ref{eq:app:M1}) allow for diagrammatic calculation of $\mathcal{M}_0(\mathbf{k})$ and $\mathcal{M}_1(\mathbf{k})$. To do that, we approximate the true ground state by the variational solution  $\left.P_G |\Psi_0\right>$.  Explicitly, 

\begin{eqnarray}
  &\mathcal{M}_0(\mathbf{k}) \approx \left<\hat{n}_{\mathbf{k}\sigma}\right>_G =  \langle c^\dagger_{i\sigma}c_{i\sigma} \rangle_G + \sum \limits_{\boldsymbol{\delta} \neq \mathbf{0}} \exp(i \mathbf{k} \boldsymbol{\delta}) \langle c^\dagger_{i + \boldsymbol{\delta},\sigma} c_{i\sigma} \rangle_G = \nonumber \\ = & n_G + \sum \limits_{\boldsymbol{\delta} \neq \mathbf{0}} \exp(i \mathbf{k} \boldsymbol{\delta}) \left[ q^2 T^{11}_{i + \boldsymbol{\delta}, i} + 2 q \alpha T^{13}_{i + \boldsymbol{\delta}, i} + \alpha^2 T^{33}_{i + \boldsymbol{\delta}, i}\right] = \nonumber \\ =& n_G + q^2 T^{11}(\mathbf{k}) + 2 q \alpha T^{13}(\mathbf{k}) + \alpha^2 T^{33}(\mathbf{k}) - \nonumber\\&- \int \frac{d^d \mathbf{q}}{(2\pi)^d} \left[q^2 T^{11}(\mathbf{q}) + 2 q \alpha T^{13}(\mathbf{q}) + \alpha^2 T^{33}(\mathbf{q}\right)],
\label{eq:app:m0_diagrammatic}
\end{eqnarray}

\noindent
where $\boldsymbol{\delta}$ runs over lattice sites, $d$ is the spatial dimension, and $n_G \equiv \langle c^\dagger_{i\sigma}c_{i\sigma} \rangle_G$. The $\mathbf{k}$-space diagrammatic sums are defined as $T^{\alpha\beta}(\mathbf{k}) = \sum_{\boldsymbol{\delta}} \exp(i \mathbf{k} \boldsymbol{\delta}) \cdot T^{\alpha\beta}_{i+\boldsymbol{\delta}, i}$ (cf. \ref{appendix:graphs}). The last term in equation~(\ref{eq:app:m0_diagrammatic}) arises from the summation restriction $\boldsymbol{\delta} \neq \mathbf{0}$. By performing a similar calculation for the first moment, one arrives at

\begin{eqnarray}
  &\mathcal{M}_1(\mathbf{k}) \approx  -\langle c_\mathbf{k \sigma}^\dagger [\mathcal{H} - \mu \hat{N}, c_\mathbf{k \sigma}] \rangle_G = \nonumber \\ = &  (\epsilon_\mathbf{k} - \mu) \cdot \langle \hat{n}_{\mathbf{k}\sigma}\rangle_G + U \cdot \sum\limits_{\boldsymbol{\delta}} \mathrm{e}^{i \mathbf{k} \boldsymbol{\delta}} \langle c^\dagger_{i+\boldsymbol{\delta}, \sigma} \hat{n}_{i \bar{\sigma}} c_{i\sigma}\rangle_G.
\end{eqnarray}

\noindent
By using the relation $P_{G i} \hat{n}_{i\bar{\sigma}}c_{i\sigma} P_{G i} = q' c_{i\sigma} + \alpha' \hat{n}^\mathrm{HF}_{i\bar{\sigma}}c_{i\sigma} $ with $q' = \lambda_{\bar{\sigma}} \lambda_{\uparrow\downarrow} n^0_{\bar{\sigma}}$ and $\alpha' = \lambda_{\bar{\sigma}} \lambda_{\uparrow\downarrow}$, we get

\begin{samepage}
\begin{eqnarray}
  \mathcal{M}_1(\mathbf{k}) =&  (\epsilon_\mathbf{k} - \mu) \cdot n_G + U d^2_G + [(\epsilon_\mathbf{k} - \mu) \cdot q^2 + U q q'] \times \nonumber\\&\times \left(T^{11}(\mathbf{k}) - \int \frac{d^d\mathbf{q}}{(2\pi)^d}T^{11}(\mathbf{q})\right) + \nonumber\\ &+  [2 (\epsilon_\mathbf{k} - \mu) \cdot q\alpha + U (q \alpha' + q' \alpha)] \times  \left(T^{13}(\mathbf{k}) - \int \frac{d^d\mathbf{q}}{(2\pi)^d}T^{13}(\mathbf{q})\right) +  \nonumber\\ &+  [(\epsilon_\mathbf{k} - \mu) \cdot \alpha^2 + U \alpha \alpha'] \times  \left(T^{33}(\mathbf{k}) - \int \frac{d^d\mathbf{q}}{(2\pi)^d}T^{33}(\mathbf{q})\right),
\end{eqnarray}
\end{samepage}

\noindent
where $d_G^2 \equiv \langle\hat{n}_{i\uparrow}\hat{n}_{i\downarrow}\rangle_G$ is the probability of double orbital occupancy. The remaining task is to diagrammatically compute the $T^{11}(\mathbf{k})$, $T^{13}(\mathbf{k})$, and $T^{33}(\mathbf{k})$ contributions, which is detailed in \ref{appendix:graphs}.

To illustrate the general scheme, developed in this work, we write an explicit expression for $n_\mathbf{k}$ in the paramagnetic state up to the first-order order $k$ (note that in actual calculations at least second-order terms should be included to obtain reliable result). This is done with the use of equation~(\ref{eq:app:m0_diagrammatic}) and the diagrammatic decomposition described in \ref{appendix:graphs} (cf. also figure~\ref{fig:diagrams}). For two-dimensional lattice ($d = 2$), one obtains

\begin{eqnarray}
  \label{eq:explicit_expression_nk_1st_order}
  & n_\mathbf{k}  \approx  n_G + q^2\tilde{P}_\mathbf{k} + 2 q x \alpha \int \frac{d^2\mathbf{q}_0}{(2\pi)^2} \frac{d^2\mathbf{q}_1}{(2\pi)^2} \tilde{P}_\mathbf{k} \tilde{P}_{\mathbf{k} + \mathbf{q}_0 + \mathbf{q}_1}\tilde{P}_{-\mathbf{q}_0} \tilde{P}_{-\mathbf{q}_1} + \nonumber \\ & - \alpha^2 \int \frac{d^2\mathbf{q}_0}{(2\pi)^2} \frac{d^2\mathbf{q}_0}{(2\pi)^2} \tilde{P}_{\mathbf{k} + \mathbf{q}_0 + \mathbf{q}_1}\tilde{P}_{-\mathbf{q}_0} \tilde{P}_{-\mathbf{q}_1} + \nonumber \\ & - 2 \alpha^2 x \int \frac{d^2\mathbf{q}_0}{(2\pi)^2} \frac{d^2\mathbf{q}_1}{(2\pi)^2} \frac{d^2\mathbf{q}_2}{(2\pi)^2} \tilde{P}_{-\mathbf{k}_0 - \mathbf{k}_1 - \mathbf{k}_2} \tilde{P}_{\mathbf{k}_2} \tilde{P}_{-\mathbf{k}_0} \tilde{P}_{-\mathbf{k}_1} \tilde{P}_{\mathbf{k} + \mathbf{k}_0 + \mathbf{k}_1} - \nonumber \\ & - q^2 \int \frac{d^2\mathbf{k}}{(2\pi)^2} \tilde{P}_\mathbf{k} - 2 q \alpha x \int \frac{d^2\mathbf{k}}{(2\pi)^2} \frac{d^2\mathbf{q}_0}{(2\pi)^2} \frac{d^2\mathbf{q}_1}{(2\pi)^2} \tilde{P}_{\mathbf{k} + \mathbf{q}_0 + \mathbf{q}_1}\tilde{P}_{-\mathbf{q}_0} \tilde{P}_{-\mathbf{q}_1} + \nonumber \\ & + \alpha^2 \int \frac{d^2\mathbf{k}}{(2\pi)^2} \frac{d^2\mathbf{q}_0}{(2\pi)^2} \frac{d^2\mathbf{q}_0}{(2\pi)^2} \tilde{P}_{\mathbf{k} + \mathbf{q}_0 + \mathbf{q}_1}\tilde{P}_{-\mathbf{q}_0} \tilde{P}_{-\mathbf{q}_1} + \nonumber \\ & + 2 \alpha^2 x \int \frac{d^2\mathbf{k}}{(2\pi)^2} \frac{d^2\mathbf{q}_0}{(2\pi)^2} \frac{d^2\mathbf{q}_1}{(2\pi)^2} \frac{d^2\mathbf{q}_2}{(2\pi)^2} \tilde{P}_{-\mathbf{k}_0 - \mathbf{k}_1 - \mathbf{k}_2} \tilde{P}_{\mathbf{k}_2} \tilde{P}_{-\mathbf{k}_0} \tilde{P}_{-\mathbf{k}_1} \tilde{P}_{\mathbf{k} + \mathbf{k}_0 + \mathbf{k}_1},
\end{eqnarray}

\noindent
where $\tilde{P}_\mathbf{k}$ is the shifted paramagnetic line, introduced in \ref{appendix:graphs}, and $x$ denotes correlator parameter. Note that at zeroth order $n_G = n_\sigma^0$ and $n_\mathbf{k} = (1 - q^2) n^0_\sigma + q^2 n^0_\mathbf{k}$, where $n^0_\mathbf{k}$ is the distribution function for he Fermi sea of non-interacting quasiparticles. In the strong-correlation limit, where $q^2 \ll 1$, the distribution function is almost flat across the Brillouin zone with only small discontinuity at the Fermi level due to loss of quasiparticle coherence. The latter behavior also provides the basis for the distinction between itinerant ($\propto q^2$) and localized ($\propto 1 - q^2$) contributions to physical quantities \cite{SpalekPhysRevB1986}.

\section{Real-space cutoff scaling of the DE-GWF method: The case of local quantities}
\label{appendix:real_space_scaling}

\begin{figure*}
\includegraphics[width = 1.0\textwidth]{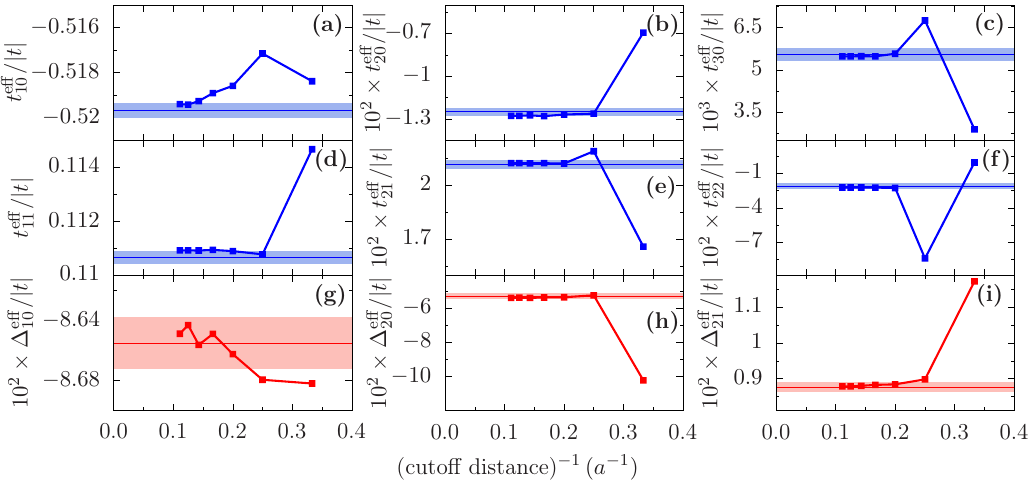} %
\caption{Scaling of selected effective Hamiltonian parameters with the real-space correlation-range cutoff for the DE-GWF method (connected squares). Vertical blue lines and shaded regions show corresponding $\mathbf{k}$-DE-GWF values along with statistical uncertainties, obtained without the cutoff. The model parameters were set to $t'/|t| = 0.25$, $U/|t| = 20$,  $J = 1/3$, and $\delta \approx 0.20$. Panels (a)-(f) show the effective hopping integrals, whereas in panels (g)-(i) SC gap parameters are displayed.}%
  \label{fig:gwf_scaling}
\end{figure*}

The real-space DE-GWF method is not suitable for description of wave-vector-resolved quantities, but it is expected to be robust in regard of local correlations (by local we mean those involving operators acting on lattice sites separated by up to several lattice constants). To verify this conjecture, as well as to test of our implementations of both DE-GWF and $\mathbf{k}$-DE-GWF algorithms, we have performed a systematic analysis of finite-cutoff-range effects on the DE-GWF parameters of effective Hamiltonian $\mathcal{H}_\mathrm{eff}$ and compared them with the thermodynamic-limit $\mathbf{k}$-DE-GWF values. In figure~\ref{fig:gwf_scaling} we plot selected effective hopping integrals (blue squares) and SC gap parameters (red squares) as a function of inverse cutoff range (measured in $a^{-1}$ with $a$ being the lattice spacing). The lines connecting points are guides to the eye. The calculations have been performed only to the second expansion order ($k \leq 2$) so that real-space cutoffs larger than usual $3$-$5$ lattice constants could be considered. The solid horizontal lines are the $\mathbf{k}$-DE-GWF values with statistical uncertainties marked by shaded regions. For the largest real-space cutoffs, we get an excellent agreement. Note that the variation of  effective parameters in the $3$-$10$ lattice-constant cutoff range exceeds the statistical uncertainties of the $\mathbf{k}$-DE-GWF result in some cases. This confirms the usefulness of the $\mathbf{k}$-DE-GWF approach to study local quantities as well.

Finally, in table \ref{table:gwf_comparison} we present detailed comparison of the DE-GWF and $\mathbf{k}$-DE-GWF results for the $t$-$J$-$U$ model with $t/t' = -0.25$, $U/|t| = 20$, $J/|t| = 1/3$, and hole-doping $\delta = 0.198196$. The cutoff of $8$ lattice sites has been taken in the DE-GWF calculation. Note the excellent agreement between the two approaches for the real-space quantities.

\begin{table}
  \caption{Comparison of the variational parameters obtained within the real-space DE-GWF and reciprocal-space  $\mathbf{k}$-DE-GWF techniques for the $t$-$J$-$U$ model with $t/t' = -0.25$, $U/|t| = 20$, $J/|t| = 1/3$, and hole-doping $\delta = 0.198196$. In both cases the calculations have been performed up to the second order of diagrammatic expansion ($k \leq 2$). The real-space cutoff for the DE-GWF method has been set to 9 lattice sites, whereas the $\mathbf{k}$-space sampling was set to $16 \cdot 10^7$ within the $\mathbf{k}$-DE-GWF approach. For the $\mathbf{k}$-DE-GWF data statistical uncertainties are given if available. Here $x$ is the correlator parameter, $E_G$ denotes ground-state energy, $\mu$ is the chemical potential, $d^2$ is the probability of double site occupancy,  $n_0 \equiv \langle\Psi_0|\hat{n}_{i\sigma}|\Psi_0\rangle$, and $n_0 \equiv \langle\Psi_G|\hat{n}_{i\sigma}|\Psi_G\rangle/\langle\Psi_G|\Psi_G\rangle$. The remaining quantities marked by the superscript ``eff'' are the parameters of the effective Hamiltonian $\mathcal{H}_\mathrm{eff}$.} \label{table:gwf_comparison}
  \centering
  \vspace{1em}
  \begin{tabular}{ c c c c c c}
    \hline\hline
   \multicolumn{1}{c}{Parameter} & \multicolumn{1}{c}{DE-GWF} & \multicolumn{1}{c}{$\mathbf{k}$-DE-GWF} & \multicolumn{1}{c}{Parameter} & \multicolumn{1}{c}{DE-GWF} & \multicolumn{1}{c}{$\mathbf{k}$-DE-GWF} \\
    \hline\hline
  $t^\mathrm{eff}_{10}/|t|$ & $-0.51940$ & $-0.51968 \pm 0.00035$ &  $\Delta^\mathrm{eff}_{10}/|t|$ & $-0.08649$ & $-0.08655 \pm 0.00018$  \\
  $t^\mathrm{eff}_{11}/|t|$ & $0.11093$ & $0.11071 \pm 0.00025$ &   $\Delta^\mathrm{eff}_{20}/|t|$ & $-0.00541$ & $-0.00533 \pm 0.00017$ \\
  $t^\mathrm{eff}_{20}/|t|$ & $-0.01279$ & $-0.01251 \pm 0.00026$ &  $\Delta^\mathrm{eff}_{21}/|t|$ & $0.00878$ & $0.00876 \pm 0.00015$ \\
  $t^\mathrm{eff}_{21}/|t|$ & $0.02123$ & $0.02103 \pm 0.00028$ &  $\Delta^\mathrm{eff}_{30}/|t|$ & $0.00368$ & $0.00374 \pm 0.00015$ \\
  $t^\mathrm{eff}_{22}/|t|$ & $-0.00222$ & $-0.00211 \pm 0.00026$ &  $\Delta^\mathrm{eff}_{31}/|t|$ & $-0.00036$ & $-0.00038 \pm 0.00016$ \\
  $t^\mathrm{eff}_{30}/|t|$ & $0.00548$ & $0.00554 \pm 0.00025$ &  $\Delta^\mathrm{eff}_{32}/|t|$ & $-0.00016$ & $-0.00025 \pm 0.00014$ \\
  $t^\mathrm{eff}_{31}/|t|$ & $-0.00008$ & $-0.00008 \pm 0.00026$ & $\mu^\mathrm{eff}/|t|$ & $-0.53049$ & $-0.52713$  \\
  $t^\mathrm{eff}_{32}/|t|$ & $-0.00193$ & $-0.00191 \pm 0.00026$ &  $x$ & $-2.58255$ & $-2.58319$ \\
  $E_G/|t|$ & $-0.649981$ & $-0.65004 \pm 0.00046$ &  $d^2$ & $0.005763$ & $0.005762 \pm 0.000008$ \\
  $\mu/|t|$ & $1.4047$ & $1.40907$ &  $n_0$ & $0.392884$ & $0.392956$ \\
    $n_G$ & $0.400902$ & $0.400903 \pm 0.000026$ &  $$ & $$ & $$ \\
   \hline\hline                                                                                        
\end{tabular}%
\end{table}

\section{Spectral function moments for approximate ground states}
\label{appendix:artifacts}

Here we address the consequences of evaluating the spectral function moments using the approximate ground state, the special case of which is the variational solution. Specifically, we show that using the exact relation $\mathcal{M}_1(\mathbf{k}) = -\langle \Psi_\mathrm{GS}| c^\dagger_{\mathbf{k}\sigma} [\mathcal{H} - \mu \hat{N}, c_{\mathbf{k}\sigma}]|\Psi_\mathrm{GS}\rangle$, but with the approximate ground state $|\Psi_\mathrm{approx}\rangle$ instead of the exact one $|\Psi_\mathrm{GS}\rangle$, generically induces artifacts in the form of $\mathcal{M}_1(\mathbf{k})$. This can be illustrated by referring to the non-interacting and thus exactly diagonalizable model, given by the Hamiltonian $\mathcal{H}_\mathrm{0} = \sum_{\mathbf{k}\sigma} \epsilon_\mathbf{k}^\mathrm{exact} c^\dagger_{\mathbf{k}\sigma} c_{\mathbf{k}\sigma}$, where $\epsilon_\mathbf{k}^\mathrm{exact}$ is the exact quasiparticle dispersion. We fix the parameters by taking $\epsilon_\mathbf{k}^\mathrm{exact} = -2 |t| (\cos k_x + \cos k_y) + 4 \cdot 0.25 |t| \cos k_x \cos k_y$, where $|t|$ is the magnitude of the nearest-neighbor hopping. Additionally, we set the doping level at $\delta = 0.1$, corresponding to the chemical potential $\mu \approx -0.8046 |t|$. Now, instead of using $|\Psi_\mathrm{GS}\rangle$ while calculating the quasiparticle velocities, we introduce the approximate wave function $|\Psi_\mathrm{approx}\rangle$ as the ground state of a slightly modified  Hamiltonian $\mathcal{H}_0' = \sum_{\mathbf{k}\sigma} \epsilon_\mathbf{k}' c^\dagger_{\mathbf{k}\sigma} c_{\mathbf{k}\sigma}$, where $\epsilon_\mathbf{k}' = -2 \cdot 1.1 \cdot |t| (\cos k_x + \cos k_y) + 4 \cdot 0.25 |t| \cos k_x \cos k_y$ (note the factor $1.1$ multiplying the nearest-neighbor hopping term). The chemical potential has been adjusted accordingly to $\mu_0' \approx -0.8304 |t|$ to preserve the doping level. The function $|\Psi_\mathrm{approx}\rangle$ is now clearly not the ground state of $\mathcal{H}_0$, but it can be considered a reasonable approximation as it optimizes only slightly modified Hamiltonian $\mathcal{H}_0'$.

{
  
\begin{figure}
 \centering
\includegraphics[width = 0.6\columnwidth]{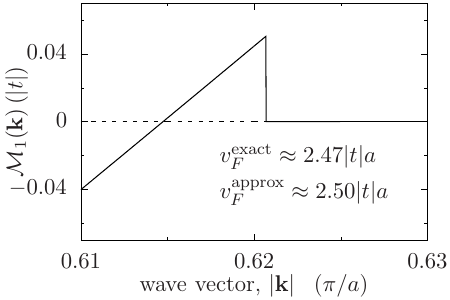} %
\caption{Illustration of the artifacts in the first spectral function moment $\mathcal{M}_1(\mathbf{k})$ (solid line) that emerge as a consequence of using an approximate ground state instead of the exact one: (i) a small jump-discontinuity of $\mathcal{M}_1(\mathbf{k})$, and (ii) Positive values of $\mathcal{M}_1(\mathbf{k})$ for the range of wave vectors. Inside the plot, the exact and approximate Fermi velocities are given ($v_F^\mathrm{exact}$ and $v^\mathrm{approx}_F$, respectively). The value of $v^\mathrm{approx}_F$ has been obtained from the discontinuity of the first wave vector derivative of $\mathcal{M}_1(\mathbf{k})$ as is described in the text. The model definition and the values of parameters, for which this plot has been generated, are detailed in \ref{appendix:artifacts}. Horizontal dashed line is guide to the eye.}%
  \label{fig:artifact}
\end{figure}
}

In figure~\ref{fig:artifact} we plot the first moment of the spectral function $\mathcal{M}_1(\mathbf{k})$ along the interval on the $\Gamma$-$M$ line encompassing the Fermi wave vector, calculated using the $|\Psi_\mathrm{approx}\rangle$ in place of $|\Psi_\mathrm{GS}\rangle$. A small discontinuity of $\mathcal{M}_1(\mathbf{k})$ is seen. Additionally, in a narrow range of wave vectors, the first moment becomes positive. These are the artifacts pointed out in the main text.

To check first the impact of the approximate ground state on the value of calculated Fermi velocity, we have computed its exact value $v_F^\mathrm{exact} \approx 2.47 |t| a$ and contrasted it with $v_F^\mathrm{approx} \approx 2.50 |t| a$, obtained from $\mathcal{M}_1(\mathbf{k})$ of figure~\ref{fig:artifact} by the methods detailed in the main text. These two numbers differ by $\sim 1\%$. Given that the hopping has been scaled by $10\%$ in the Hamiltonian $\mathcal{H}_0'$, defining $|\Psi_\mathrm{approx}\rangle$, the method of extracting quasiparticle velocities from $\mathcal{M}_\mathbf{k}$ is robust to the use of approximate solutions in the present case.

Parenthetically, for the particular situation considered in this Appendix, the velocity obtained from the jump of $\mathcal{M}_1(\mathbf{k})$ is equal to the slope of the  \textit{exact} quasiparticle dispersion, but measured slightly away from the Fermi surface. This supports the view that singularities of $\mathcal{M}_1(\mathbf{k})$ might reflect the quasiparticle properties at energies shifted with respect to the Fermi level if the approximate ground state is used. The latter observation is compatible with the analysis performed in the main text. For generic case which cannot be studied analytically, this statement can be also justified by modeling the impact of the non-zero admixture of excited states in the approximate wave function. The simplest way to achieve this goal is to allow for the non-zero upper limit in the expression for $\mathcal{M_1(\mathbf{k})}$, i.e., define $\mathcal{M_1^\mathrm{approx}(\mathbf{k})} = \int_{-\infty}^{\Delta E} \omega \mathcal{A}(\mathbf{k}, \omega) d\omega$, where $\Delta E$ models the excited state contribution. Here we also postulate that using the approximate ground state weakly influences the spectral function itself (which should hold for sufficiently huge variational space), i.e.,  $\mathcal{A}(\mathbf{k}, \omega) \approx Z \delta(\omega - \epsilon^\mathrm{exact}_\mathbf{k}) + \mathcal{A}_\mathrm{inc}(\mathbf{k}, \omega)$ with incoherent part $\mathcal{A}_\mathrm{inc}(\mathbf{k}, \omega)$. By performing integration, one obtains $\mathcal{M}_1^\mathrm{approx}(\mathbf{k}) \approx Z \epsilon_\mathbf{k}^\mathrm{exact} \Theta(\Delta E - \epsilon_\mathbf{k}^\mathrm{exact}) + \mathrm{smooth \,\, part}$, where $\Theta$ denotes Heaviside step function. The discontinuity of the first spectral moment thus occurs at some effective Fermi wave vector $k_F$ (different from the exact value $k_F^\mathrm{exact}$) and is  approximately equal to the energy scale of the excited states admixed to the wave function: $\Delta E \approx \Delta \mathcal{M}_1 / Z$. By approximating $\epsilon^\mathrm{exact}_\mathbf{k} \approx v_F^\mathrm{extact} (k - k_F^\mathrm{exact})$, the mismatch $\Delta k_0 \equiv k_F - k_F^\mathrm{exact}$ can be quantified by noting that $\Delta E \approx v_F^\mathrm{exact} \Delta k_0$ or, equivalently, $\Delta k_0 \approx \Delta \mathcal{M}_1/(Z v_F)$.

Finally, we point out that, contrary to $\mathcal{M}_1(\mathbf{k})$, the zeroth moment $\mathcal{M}_0(\mathbf{k})$ is free of similar artifacts, even for approximate ground states. The reason is that $\mathcal{M}_0(\mathbf{k})$ is equal to the statistical distribution function $n_\mathbf{k}$ that fulfills $0 \leq n_\mathbf{k} \leq 1$ for any wave function.

\section*{References}
\bibliographystyle{unsrt}
\bibliography{bibliography}

\end{document}